\documentclass[twocolumn,amssymb,amsmath,aps,pra,nofootinbib]{revtex4-1}
\usepackage{graphicx}
\usepackage{multirow}
\usepackage{array}
\newcommand{\gv}{$g$-value}

\newcommand{\nucb}{\bar{\nu}_c}
\newcommand{\nuzb}{\bar{\nu}_z}
\newcommand{\numb}{\bar{\nu}_m}
\newcommand{\nuab}{\bar{\nu}_a}
\newcommand{\wcb}{\bar{\omega}_c}

\newcommand{\wmb}{\bar{\omega}_m}
\newcommand{\fcb}{\bar{f}_c}

\newcommand{\gz}{\gamma_z}
\newcommand{\gc}{\gamma_c}
\newcommand{\gp}{\gamma^\prime}

\newcommand{\Dw}{\Delta\omega}

\newcommand{\Dwc}{\Delta\omega_c}

\newcommand{\zhat}{\mathbf{\hat{z}}}
\newcommand{\rhohat}{\boldsymbol{\hat{\rho}}}

\newcommand{\ket}[1]{\left|#1\right\rangle}

\def\abs#1{\left| #1 \right|}

\newcommand{\w}{3.25in}

\begin{document}

\newcommand{\gComparisonFigure}{
\begin{figure}[htbp!]
    \centering
    \includegraphics[width=\columnwidth]{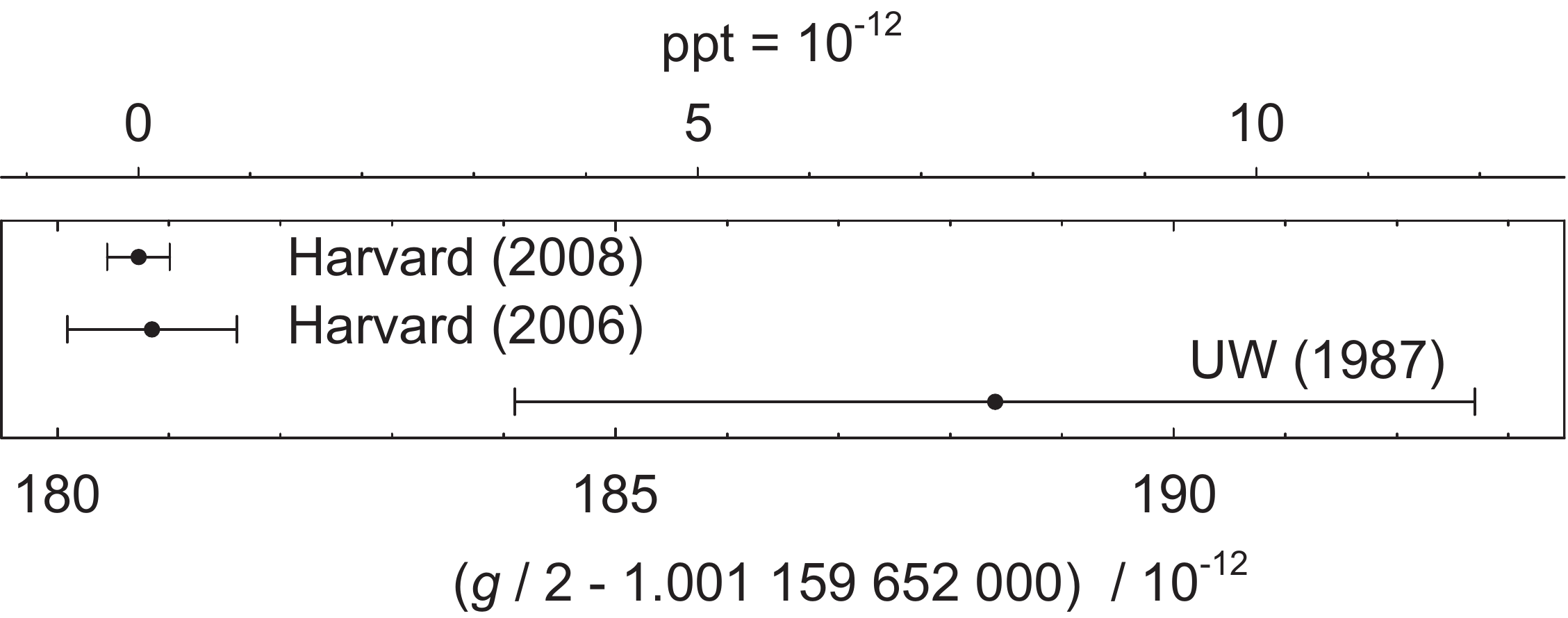}
    \caption{Measurements  \cite{HarvardMagneticMoment2008,HarvardMagneticMoment2006,DehmeltMagneticMoment} of the dimensionless magnetic moment of the electron, $g/2$, which is the electron magnetic
    moment in Bohr magnetons.}
    \label{fig:gComparison}
\end{figure}
}

\newcommand{\AlphaComparisonFigure}{
\begin{figure}[htbp!]
\centering
\includegraphics[width=\columnwidth]{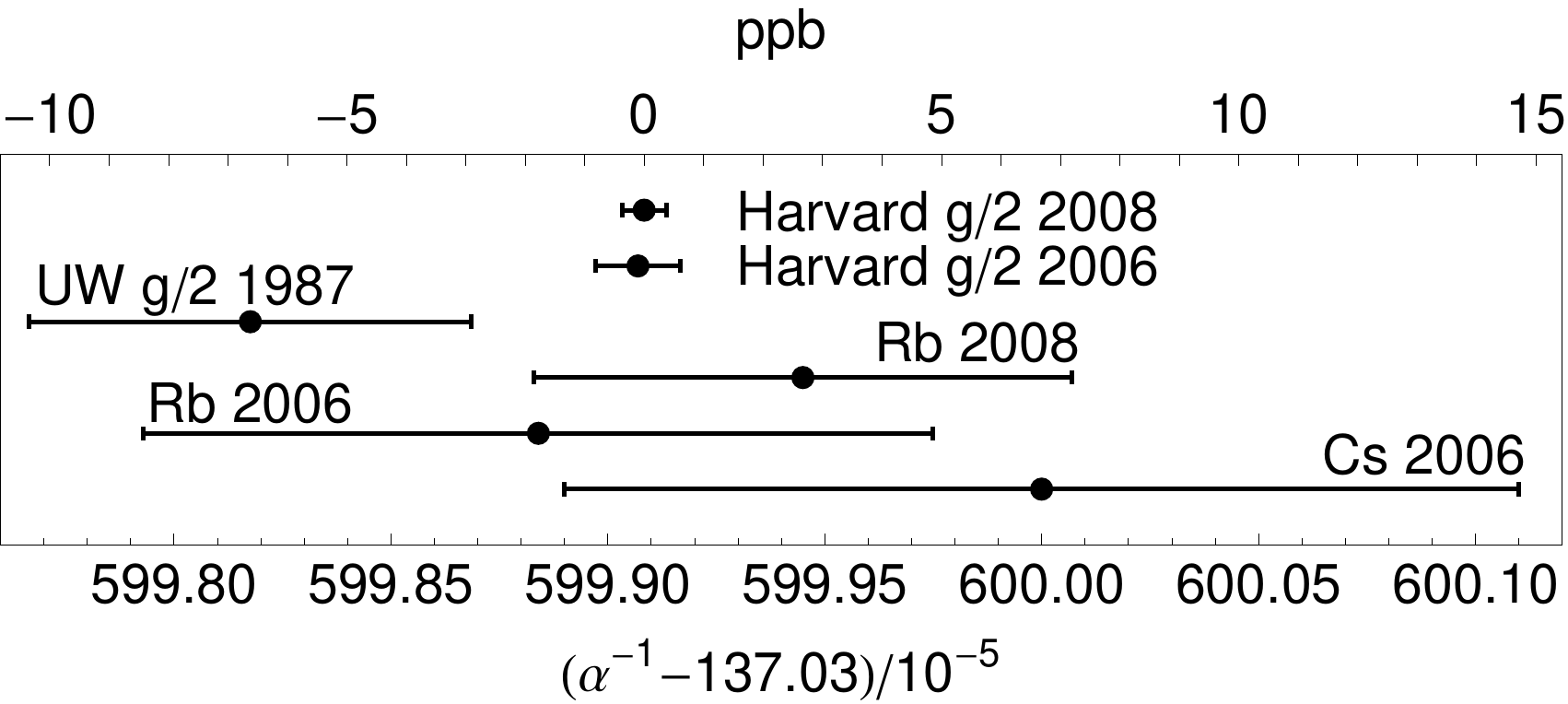}
\caption{The most precise $\alpha$ determinations~\cite{HarvardMagneticMoment2008,HarvardMagneticMoment2006,AlphaRbPRL2006,Tanner2006,AlphaRb2008}.} \label{fig:AlphaComparison}
\end{figure}
}

\title{Cavity Control of a Single-Electron Quantum Cyclotron:\\Measuring the Electron Magnetic Moment}

% The address
\newcommand{\HU}{Department of Physics, Harvard University, Cambridge, Massachusetts 02138, USA}

\author{D. Hanneke}\altaffiliation{Current address: NIST Boulder, CO 80305}
\affiliation{\HU}
\author{S. Fogwell Hoogerheide}
\affiliation{\HU}
\author{G. Gabrielse}
\email[E-mail: ]{gabrielse@physics.harvard.edu} \affiliation{\HU}

\begin{abstract}
Measurements with a one-electron quantum cyclotron determine the electron magnetic moment, given by $g/2 =
1.001\,159\,652\,180\,73\,(28)\,[0.28~\textrm{ppt}]$, and the fine structure constant, $\alpha^{-1}=137.035\,999\,084\,(51)\,[0.37~\textrm{ppb}]$.
Brief announcements of these measurements \cite{HarvardMagneticMoment2006,HarvardMagneticMoment2008} are supplemented here with a more complete
description of the one-electron quantum cyclotron and the new measurement methods, a discussion of the cavity control of the radiation field, a summary of the analysis of the measurements, and a fuller discussion of the uncertainties.
\end{abstract}

\date{Submitted to Phys.\ Rev.\ A on 3 Sept.\ 2010}

\maketitle

\section{Introduction}

\subsection{The Electron Magnetic Moment}

Measurements of the electron magnetic moment $\boldsymbol{\mu}$ probe the interaction of the electron with the fluctuating vacuum, allow the highest
accuracy determination of the fine structure constant, and sensitively test quantum electrodynamics (QED). For an eigenstate of spin $\mathbf{S}$,
\begin{equation}
    \boldsymbol{\mu}=-\frac{g}{2}\mu_B\frac{\mathbf{S}}{\hbar/2},
    \label{eq:gDef}
\end{equation}
where $g/2$ is the magnitude of $\boldsymbol{\mu}$ scaled by the Bohr magneton $\mu_B = e\hbar/(2m)$.

For angular momentum arising from orbital
motion, $g/2$ depends on the relative distribution of charge and mass and equals 1/2 if they coincide, for example cyclotron motion in a magnetic
field. For a point particle in a renormalizable Dirac description, $g/2=1$, and deviations from this value probe a particle's interactions with the
vacuum as well as the nature of the particle itself, as with the proton whose $g/2 \approx 2.8$ arises from its quark--gluon composition.

\subsection{New Measurements of the Electron Moment}

Our new measurements, announced in 2006 \cite{HarvardMagneticMoment2006} and 2008 \cite{HarvardMagneticMoment2008}, used a one-electron quantum cyclotron
\cite{QuantumCyclotron} to determine the electron $g/2$ to a $0.76$ ppt and then to a $0.28$ ppt accuracy. The latter result,
\begin{equation}
g/2 = 1.001 \, 159 \, 652 \, 180 \, 73 \, (28)~~~~~[0.28~\rm{ppt}], \label{eq:gHarvard2008}
\end{equation}
has an uncertainty that is 2.7 and 15 times smaller than the 2006 and 1987 measurements (Fig.~\ref{fig:gComparison}), the latter being a measurement
that stood for nearly twenty years \cite{DehmeltMagneticMoment}.  The electron $g$ is measured with an uncertainty that is $2300$ times smaller
than has been achieved for the heavier muon lepton \cite{gMuon2006}.

\gComparisonFigure

\newcommand{\noskip}{\\[-0.6cm]}

The central feature of the new measurements is the quantum jump spectroscopy of completely resolved cyclotron and spin levels of a one-electron
quantum cyclotron \cite{QuantumCyclotron}.  A number of new methods were introduced to make this possible.
\begin{enumerate}
\item A cylindrical Penning trap cavity that was invented for these experiments \cite{CylindricalPenningTrap} imposes boundary conditions upon the radiation
field as well as providing an electrostatic quadrupole potential in which a single particle can be suspended and observed
\cite{CylindricalPenningTrapDemonstrated}.\noskip
\item The resulting cavity-inhibited spontaneous emission, at a rate 10 to 50 times below the radiation rate in
free space, gives the averaging time required to resolve one-quantum transitions that are made when all detection systems are turned off. \noskip
\item Stored electron plasmas \cite{SynchronizedElectronsPRL,SynchronizedElectronsPRA,CavityShiftsQEDBook} and the damping of a single electron in this cavity
\cite{HarvardMagneticMoment2008} are used together to determine cavity frequency shifts and eliminate cavity shifts as a major uncertainty
\cite{HarvardMagneticMoment2008}. \noskip
\item Blackbody photons that would cause unwanted quantum jumps are eliminated by lowering the cavity temperature to 100
mK with a dilution refrigerator \cite{QuantumCyclotron}. \noskip
\item Quantum nondemolition measurements of the cyclotron and spin energy level are realized using
a one-particle self-excited oscillator \cite{SelfExcitedOscillator}. \noskip
\item The stored electron serves as its own magnetometer, allowing the accumulation of
lineshape statistics over days, revealing that a broadening of the expected lineshapes is the major remaining uncertainty
\cite{HarvardMagneticMoment2008}.\noskip
\end{enumerate}
Following sections will discuss the quantum cyclotron and the new methods.

\subsection{A Long History}

As befits one of the few properties of the electron that can be accurately measured, the Harvard magnetic moment measurements detailed here are only
the latest in a long history of measurements that make use of different methods. The early history~\cite{Rich} established that $g/2 \approx 1$. A
series of more precise measurements followed at the University of Michigan, by measuring the difference of the cyclotron and spin precession frequencies of keV
electrons traveling on helical orbits in a magnetic field, concluding with a $3500$ ppt measurement of $g/2$
\cite{Rich}. (Here ppt refers to $1$ part in $10^{12}$, and ppb refers to $1$ part in $10^{9}$.) Research groups at the University of Mainz and the University of Washington (UW) next developed methods to measure the electron magnetic moment using a large number of electrons stored in a Penning trap
\cite{GraffMeasuringCyclotronAndSpinResonancesForElectrons1968,DehmeltWalls1968,GraffAnomalyResonanceForElectrons1969,WallsSteinAnomalyForElectrons1973}.
Out of these efforts came the capability to suspend and detect a single electron in a Penning trap \cite{FirstSingleElectron1973} at the UW. A few years later a
measurement was made with one electron \cite{ContinuousSternGerlach1977}.  Over the next decade these methods were refined, culminating in the
celebrated 1987 measurement already mentioned \cite{DehmeltMagneticMoment} that reported an uncertainty of $4$ ppt.

\subsection{The Fine Structure Constant}

The fine structure constant,
\begin{equation}
    \alpha = \frac{e^2}{4\pi\epsilon_0 \hbar c},
    \label{eq:alphaintro}
\end{equation}
gives the strength of the electromagnetic coupling in the low-energy limit.  The energy scales for atoms are set by powers of $\alpha$ times the
electron rest energy, $mc^2$.  For hydrogen atoms the binding energy scale is $\alpha^2mc^2$, the fine structure splitting scale is $\alpha^4mc^2$,
and Lamb shift scale is $\alpha^5mc^2$.  The quantum Hall conductance is proportional to the fine structure constant.  The fine structure constant is
already a crucial ingredient in our system of fundamental constants~\cite{CODATA2002,CODATA2006} and it will acquire a more prominent role if plans
to redefine the SI system of units \cite{RedefineSI} go forward.

Sec.~\ref{sec:FineStructureConstant} shows how the new measurements of the electron $g/2$ determine $\alpha$ to be \cite{Alpha2006fixed,HarvardMagneticMoment2008,GabrielseAlphaChapter2009}.
\begin{eqnarray}
\alpha^{-1} &=&  137.035 \, 999 \, 084 \, (51)~~~~~~~~[0.37~\rm{ppb}].
\end{eqnarray}
The uncertainty in $\alpha$ is now limited
a bit more by the need for a higher-order QED calculation (underway~\cite{RevisedC8}) than by the measurement uncertainty in $g/2$. The total $0.37$
ppb uncertainty in $\alpha$ is now about 12 times smaller than that of the next-most-precise independent method (Fig.~\ref{fig:AlphaComparison}).

\AlphaComparisonFigure

\section{One-Electron Quantum Cyclotron}

\subsection{Electron in a Magnetic Field}

For an electron in a magnetic field, $g/2$ is specified by its spin and cyclotron frequencies, $\nu_s$ and $\nu_c$,
\begin{equation}
\frac{g}{2} = \frac{\nu_s}{\nu_c} = 1+\frac{\nu_s-\nu_c}{\nu_c} = 1+\frac{\nu_a}{\nu_c},\label{eq:gFreeSpace}
\end{equation}
or equivalently by their difference (the anomaly frequency $\nu_a \equiv \nu_s - \nu_c$) and $\nu_c$. Because $\nu_s$ and $\nu_c$ differ by only a
part-per-thousand, measuring $\nu_a$ and $\nu_c$ to a precision of 1 part in $10^{10}$ gives $g/2$ to 1 part in $10^{13}$.

Although we cannot measure accurately with one electron in free space because the electron would not stay in one place long enough, two features of
determining $g/2$ are already apparent in Eq.~\ref{eq:gFreeSpace}.  First, one can determine $g/2$ by measuring a ratio of frequencies. This is fortunate because there is nothing in physics that can be
measured more accurately than a frequency (the art of time keeping being developed being so highly developed) except for a ratio of frequencies.
Second, although both of these frequencies depend upon the magnetic field, the field dependence drops out of the ratio. The magnetic field thus needs
to be stable only on the time scale on which both frequencies can be measured, and no absolute calibration of the magnetic field is required.

\newcommand{\EnergyLevelsFigure}[1][\w]{
\begin{figure}
    \centering
    \includegraphics*[width=1.5in]{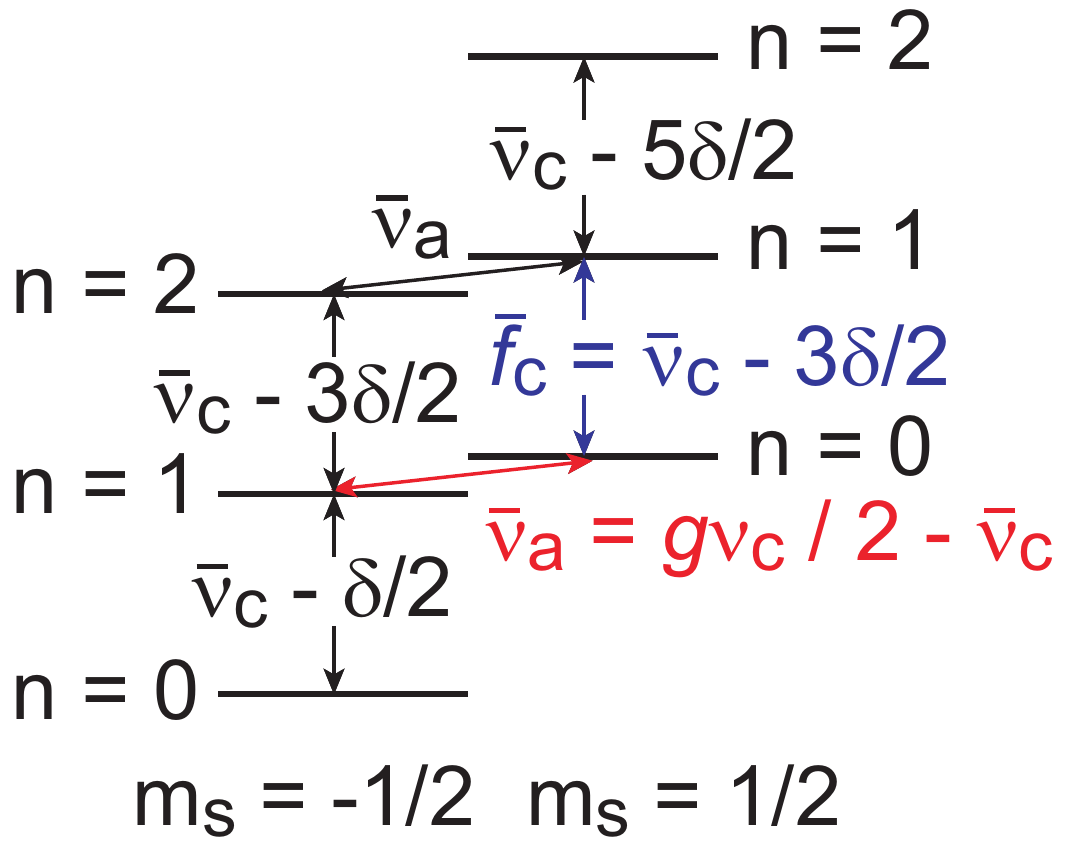}
    \caption{Lowest cyclotron and spin levels of an electron in a Penning trap.}  \label{fig:EnergyLevels}
\end{figure}
}

\newcommand{\CylindricalTrapFigure}[1][\w]{
\begin{figure}[htbp!]
\centering
\includegraphics*[width=2.5in]{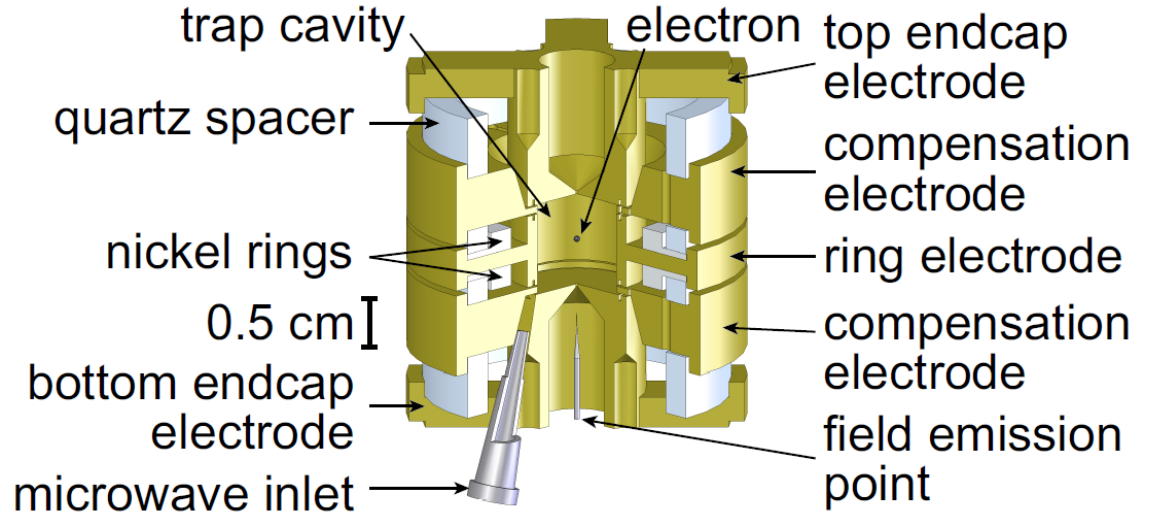}
\caption{Cylindrical Penning trap cavity used to confine a single electron and inhibit spontaneous emission.} \label{fig:CylindricalTrap}
\end{figure}
}

\subsection{Electron in a Penning Trap}

An ideal Penning trap confines an electron using a magnetic
field $B\zhat$ with an additional electrostatic quadrupole potential $V \sim z^2 - \rho^2/2$ \cite{Review}.  This potential confines the electron axially with frequency $\nuzb$ and shifts the cyclotron frequency from the free-space value $\nu_c$ to $\nucb$.  The latter frequency is
also slightly shifted by the unavoidable leading imperfections of a real laboratory trap -- a misalignment of the symmetry axis of the electrostatic
quadrupole and the magnetic field, and quadratic distortions of the electrostatic potential.

The lowest cyclotron energy levels (with quantum numbers $n=0,1,\ldots$) and the spin energy levels (with quantum numbers $m_s=\pm1/2$)
(Fig.~\ref{fig:EnergyLevels}) are given by
\begin{equation}
E(n,m_s)=\frac{g}{2} h\nu_c m_s+(n+\tfrac{1}{2})h\nucb-\tfrac{1}{2}h\delta(n+\tfrac{1}{2}+m_s)^2. \label{eq:EnergyLevels}
\end{equation}
The third term in Eq.~\ref{eq:EnergyLevels} is the leading relativistic correction \cite{Review} to the energy levels.   Special relativity makes the
transition frequency between two cyclotron levels $\ket{n,m_s}\leftrightarrow\ket{n+1,m_s}$ decrease from $\nucb$ to $\nucb + \Delta \nucb$, with the
shift
\begin{equation}
    \Delta \nucb = -\delta(n+1+m_s)
    \label{eq:relshift}
\end{equation}
depending upon the spin state and cyclotron state. This very small shift, with
\begin{equation}
\delta /\nu_c \equiv h\nu_c/(mc^2) \approx 10^{-9},
\end{equation}
is nonetheless significant at our precision. However, an essentially exact treatment of the relativsitic shift is possible because single quantum
transitions are resolved. The relativistic shift thus contributes no uncertainty to our measurement. This is a key advantage of the quantum cyclotron
over previous measurements systems \cite{DehmeltMagneticMoment}, in which an unknown distribution of cyclotron states was excited
\cite{VanDyckLossy}, each with a different relativistic shift.

\EnergyLevelsFigure

To determine $g/2$, we must rewrite Eq.~\ref{eq:gFreeSpace} in terms of measurable frequencies of an electron bound in the trap. The needed free-space cyclotron frequency, $\nu_c=eB/(2\pi m)$, is deduced by use of the Brown-Gabrielse invariance theorem~\cite{InvarianceTheorem},
\begin{eqnarray}
(\nu_c)^2 &=& (\nucb)^2  +  (\nuzb)^2 +  (\numb)^2.
\end{eqnarray}
The three measurable eigenfrequencies on the right include the cyclotron frequency $\nucb$ for the quantum cyclotron motion we have been discussing.
The second measurable eigenfrequency is the axial oscillation frequency $\nuzb$ for the nearly-harmonic, classical electron motion along the
direction of the magnetic field.  The third measurable eigenfrequency is the magnetron oscillation frequency for the classical magnetron motion along
the circular orbit for which the electric field for the trap and the motional magnetic field exactly cancel.

The invariance theorem applies for a perfect Penning trap, but also in the presence of the mentioned imperfection shifts of the eigenfrequencies for
a real trap.  This theorem, together with the well-defined hierarchy of trap eigenfrequencies, $\nucb\gg\nuzb\gg\numb\gg\delta$, yields an
approximate expression that is sufficient at our accuracy.  We thus determine the electron $g/2$ using
\begin{equation}
    \frac{g}{2}=\frac{\nucb+\nuab}{\nu_c}\simeq
        1+\frac{\nuab - \nuzb^2/(2\fcb)}{\fcb+3\delta/2+\nuzb^2/(2\fcb)}
        +\frac{\Delta g_{cav}}{2}.
    \label{eq:THEgEQ}
\end{equation}
The determination requires four inputs. First and second are high-precision measurements of the transition frequencies
\begin{eqnarray}
\fcb &\equiv& \nucb - \frac{3}{2}\delta\\
\nuab &\equiv& \frac{g}{2}\nu_c -\nucb
\end{eqnarray}
represented by the red and blue arrows in Fig.~\ref{eq:EnergyLevels}. Third is a relatively lower precision measurement of the axial frequency $\nuzb$. Fourth is the cavity shift $\Delta g_{cav}/2$ that arises from the interaction of the cyclotron motion and the trap cavity and is discussed in detail in Sec.~\ref{sec:CavityShifts}.

\section{Experimental Realization} \label{sec:ExperimentalRealization}

\subsection{Cylindrical Penning Trap}

A cylindrical Penning trap (Fig.~\ref{fig:CylindricalTrap}) is the key device that makes these measurements possible. It was invented
\cite{CylindricalPenningTrap} and demonstrated \cite{CylindricalPenningTrapDemonstrated} to provide boundary conditions that produce a controllable
and understandable radiation field within the trap cavity.  Spontaneous emission can be significantly inhibited at the same time as corresponding
shifts of the electron's oscillation frequencies are avoided.  The latter has not been possible \cite{Gabrielse88d} with the hyperbolic Penning traps
of earlier experiments \cite{DehmeltMagneticMoment}, which have electrodes approximating the equipotentials of an electrostatic quadrupole.

\CylindricalTrapFigure

The first function of the trap electrodes is to produce a very good approximation to an electrostatic quadrupole potential.  This requires careful choice of the relative geometry of the electrodes \cite{CylindricalPenningTrap}. The electrodes of the
cylindrical trap are symmetric under rotations about the center axis ($\zhat$), which is parallel to the spatially uniform magnetic field ($B\zhat$).
The potential (about $100$ V) applied between the endcap electrodes and the ring electrode provides the basic trapping potential and sets the axial
frequency $\nuzb$ of the nearly harmonic oscillation of the electron parallel to the magnetic field. The potential applied to the compensation
electrodes is adjusted to tune the shape of the potential, to make the oscillation as harmonic as possible. The tuning does not change $\nuzb$ very
much owing to an orthogonalization \cite{OrthogonalCompensate,CylindricalPenningTrap} that arises from the geometry choice.  What we found was that
one electron could be observed within a cylindrical Penning trap with as good or better signal-to-noise ratio than was realized in hyperbolic Penning
traps.

\newcommand{\ApparatusFigure}[1][\w]{
\begin{figure}
    \includegraphics*[height=3in]{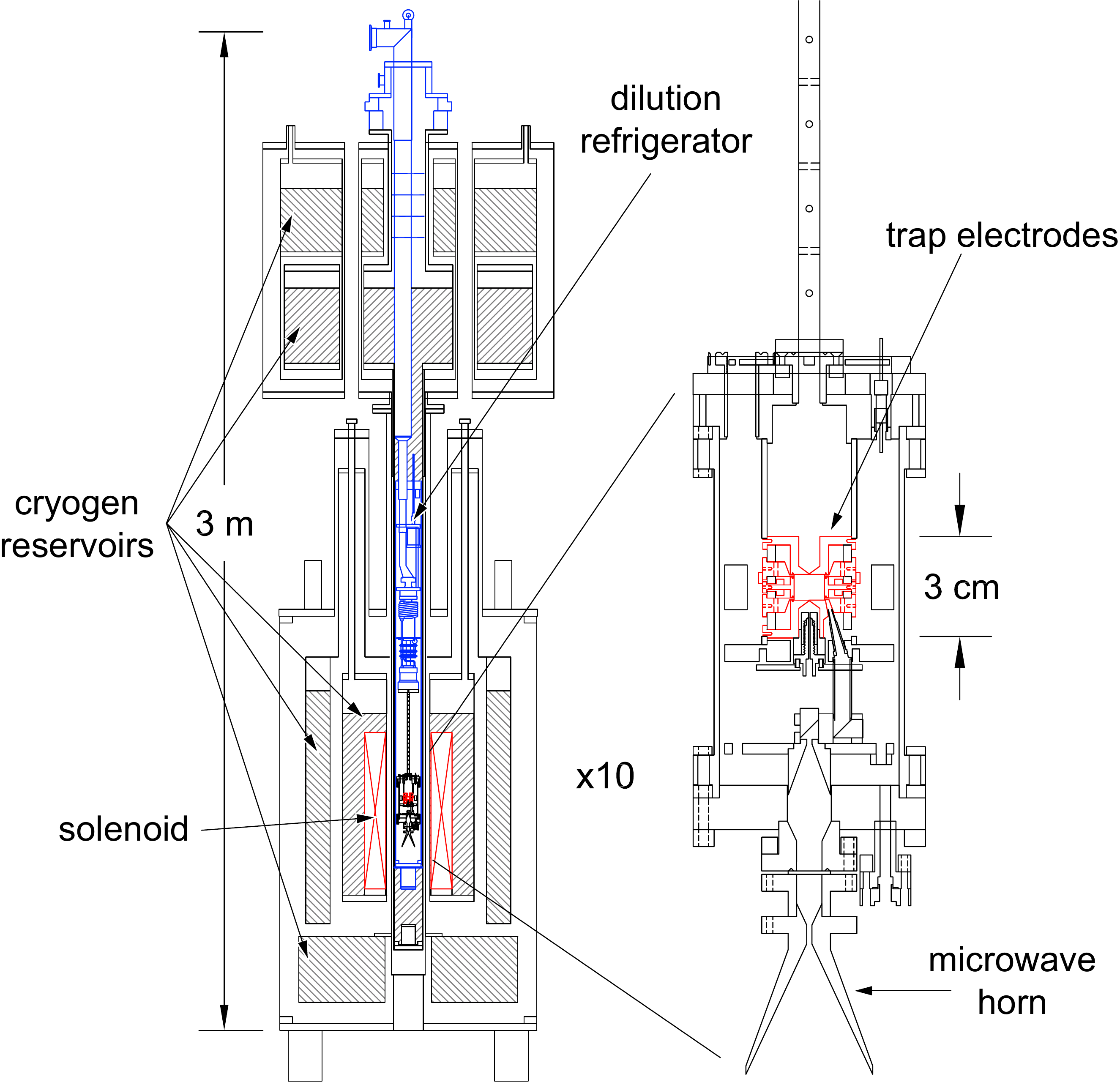}
    \caption{The apparatus. The solenoid and electrodes that form the Penning trap are in red.
    The dilution refrigerator is in blue. Cryogen spaces are hatched.}\label{fig:Apparatus}
\end{figure}
}

\subsection{Radiation Field in a Cylindrical Trap}

The second function of the trap electrodes is to form a microwave cavity whose radiation properties are well understood and controlled.  The density of states for a perfect right circular cylinder are the
familiar transverse electric (TE) and transverse magnetic (TM) radiation modes for such a geometry.  In the real trap cavity, the perturbation caused
by the small space between the electrodes is minimized by the use of ``choke flanges'' -- small channels that tend to reflect the radiation leaking
out of the trap back to cancel itself, and thus to minimize the losses from the trap. The measured radiation modes (Fig.~\ref{fig:modemap}) are close
enough to the calculated frequencies for a perfect cylindrical cavity that we have been able to identify more than 100 different radiation modes for
such trap cavities \cite{SynchronizedElectronsPRL}. The spatial properties of the electric and magnetic field for the radiation that builds up within the cavity are thus quite well
understood. Some of the modes couple to cyclotron motion of an electron centered in the cavity, others couple to the spin of a centered electron, and
still others have the symmetry that we hope will one day allow us to sideband cool the axial motion.

A right circular cylinder of diameter 2$\rho_0$ and height 2$z_0$ admits two classes of electromagnetic fields, transverse electric TE$_{mnp}$ and
transverse magnetic TM$_{mnp}$, each identified by three indices that describe their specific geometry (see e.g. \cite[Sec.\,8.7]{Jackson3rdEd}). These TE
and TM modes have characteristic frequencies,
\begin{subequations} \label{eq:modefreq}
\begin{align}
   % \wE &= c\sqrt{\left(\frac{x_{mn}^\prime}{\rho_0}\right)^2+\left(\frac{p\pi}{2 z_0}\right)^2} \\
   % \wM &= c\sqrt{\left(\frac{x_{mn}}{\rho_0}\right)^2+\left(\frac{p\pi}{2 z_0}\right)^2} ,
     \textrm{TE:}~~\omega_{mnp} &= c\sqrt{\left(\frac{x_{mn}^\prime}{\rho_0}\right)^2+\left(\frac{p\pi}{2 z_0}\right)^2} \\
    \textrm{TM:}~~\omega_{mnp} &= c\sqrt{\left(\frac{x_{mn}}{\rho_0}\right)^2+\left(\frac{p\pi}{2 z_0}\right)^2} ,
\end{align}
\end{subequations}
that are indexed with integers
%\begin{eqnarray}
%m &=& 0, 1, 2, \cdots \\
%n &=& 1, 2, 3, \cdots \\
%p &=& 1, 2, 3, \cdots,\\
%\end{eqnarray}
\begin{gather}
m = 0, 1, 2, \cdots \notag \\
n = 1, 2, \cdots \\
p = \textrm{TE:\,} 1, 2, \cdots; \textrm{TM:\,} 0, 1, \cdots \notag
\end{gather}
and are functions of the $n$th zeros of Bessel functions and their derivatives
\begin{eqnarray}
J_m(x_{mn})=0\\
J_m^\prime (x_{mn}^\prime)=0
\end{eqnarray}
The zeros force the boundary conditions at the cylindrical wall. All but the $m=0$ modes are doubly degenerate.

Of primary concern is the magnitude of the transverse electric fields since only these
components couple to cyclotron motion. For both TE and TM modes, the transverse
components of $\mathbf{E}$ are proportional to
\begin{equation}
    \sin(\tfrac{p\pi}{2}(\tfrac{z}{z_0}+1)) = \begin{cases}
                                (-1)^{p/2}\sin(\tfrac{p\pi z}{2 z_0}) & \text{for even $p$,} \\
                                (-1)^{(p-1)/2}\cos(\tfrac{p\pi z}{2 z_0}) & \text{for odd $p$.} \\
                                                                                                                                    \end{cases}
    \label{eq:transversezdependence}
\end{equation}
For an electron close to the cavity center, ($z\approx0$), only modes with odd $p$ thus have any appreciable coupling.

The transverse components of the electric fields are also proportional to either the
order-$m$ Bessel  function times $m/\rho$ or to the derivative of the
order-$m$ Bessel function. Close to the cavity center ($\rho\approx 0$),
\begin{subequations} \label{eq:besselrhoexpansion}
\begin{align}
    \frac{m}{\rho}~J_m(x_{mn}^{(\prime)}\tfrac{\rho}{\rho_0}) &\sim \begin{cases}
                        \dfrac{\rho^{m-1}}{(m-1)!}\left(\dfrac{x_{mn}^{(\prime)}}{2\rho_0}\right)^{\!m}  & \text{for}~m > 0 \\
                        0                    & \text{for}~m = 0 \end{cases} \\
    \frac{x_{mn}^{(\prime)}}{\rho_0} J_m^\prime(x_{mn}^{(\prime)}\tfrac{\rho}{\rho_0}) &\sim \begin{cases}
                        \dfrac{\rho^{m-1}}{(m-1)!}\left(\dfrac{x_{mn}^{(\prime)}}{2\rho_0}\right)^{\!m}  & \text{for}~m > 0 \\
                        -\dfrac{x_{0n}^{(\prime)2}}{2\rho_0^2} \rho          & \text{for}~m = 0 .\end{cases}
\end{align}
\end{subequations}
In the limit $\rho \rightarrow 0$, all but the $m=1$ modes vanish.

For a perfect cylindrical cavity the only radiation modes that couple to an electron
perfectly centered in  the cavity are TE$_{1n\text{(odd)}}$ and TM$_{1n\text{(odd)}}$. If
the electron is moved slightly off center axially it will begin to couple to radiation
modes with $mnp = 1n\text{(even)}$.  If the electron is moved slightly off-center radially
it similarly begins to couple to modes with $m\neq1$.

\subsection{100 mK and 5 T}

The trap cavity is cooled to 0.1 K or below via thermal contact with the mixing chamber of an Oxford Instruments Kelvinox 300 dilution refrigerator
(Fig.~\ref{fig:Apparatus}). They are housed within a separate vacuum enclosure that is entirely at the base temperature. Measurements on an apparatus
with a similar design but at 4.2~K found the vacuum in the enclosure to be better than $5\times10^{-17}$~torr~\cite{PbarMass}.  Our much lower
temperature should make our background gas pressure much lower.  We are able to keep one electron suspended in our apparatus for as long as desired --
regularly months at a time. Substantial reservoirs for liquid helium and liquid nitrogen make it possible to keep the trap cold for five to seven
days before the disruption of adding more cryogens is required.

\ApparatusFigure

The trap electrodes and their vacuum container are located within a superconducting solenoid (Fig.~\ref{fig:Apparatus}) that makes a very homogeneous magnetic
field over the interior volume of the trap cavity. A large dewar sitting on top of the solenoid dewar provides the helium needed around the dilution
refrigerator below. The superconducting solenoid is entirely self-contained, with a bore that can operate from room temperature down to 77~K. It
possesses shim coils capable of creating a field homogeneity better than a part in $10^{8}$ over a $1~\text{cm}$ diameter sphere and has a passive
``shield'' coil that reduces fluctuations in the ambient magnetic field~\cite{SelfShieldingSolenoid,SelfShieldingSolenoidDemonstrated}. When properly
energized (and after the steps described in the next section have been taken) it achieves field stability better than a part in $10^9$ per hour.  We
regularly observe drifts below $10^{-9}$ per night.

\begin{table}[b]
                \caption[Typical trap parameters and degrees of freedom]{Typical trap parameters as well as frequencies and damping rates for each degree of freedom. The damping rates include coupling to a detection circuit for $\gz$ and inhibited spontaneous emission for $\gc$. To sample the radiation modes of the electrode cavity, we change $B$, and hence $\nucb$, $\gc$, and $\nu_s$ (see Sec.~\ref{sec:CavityShifts}).}\label{tbl:trapparameters}
    \begin{ruledtabular}
      \begin{tabular}{rcrcr}
        magnetic field                                  & $B$                           & 5.36~T          && \\
        electrode potential             & $V_0$                 & 101.4~V   && \\
        electrode cavity radius & $\rho_0$  & 4.5~mm    && \\
        electrode cavity height & $2z_0$                & 7.7~mm          && \\
        \hline
        &&&& \vspace{-3mm}\\ % This prevents the exponent on the first \gamma from running into the hline
        magnetron                                                               &       $\numb$          & 133~kHz   & $\gamma_m^{-1}$ & 4~Gyr \\
                                axial                                                                           & $\nuzb$   & 200~MHz   & $\gz^{-1}$      & 0.2~s \\
                                cyclotron                                                               & $\nucb$   & 150.0~GHz & $\gc^{-1}$      & 3.7~s \\
                                spin                                                                      & $\nu_s$   & 150.2~GHz & $\gamma_s^{-1}$ & 2~yr
                        \end{tabular}
                \end{ruledtabular}
\end{table}

%\begin{tabular}{|l|c|r|}
%  \hline
%  % after \\: \hline or \cline{col1-col2} \cline{col3-col4} ...
%  Cyclotron frequency & $\omega_c/(2\pi)$ & $150$ GHz \\
%  \hline
%   Magnetic field & B & $5.3$ Tesla \\
%   Trapping potential & $V_0$ & $100$ V \\
%   Harmonic coefficient & $C_2$ & $0.5$ \\
%   Effective trapping potential & $V_0 C_2$ & $60$ V \\
%   Trap dimension & $d$ & $5$ mm\\
%  \hline
%  Trap-modified cyc.\ freq.\ & $\omega_+/(2\pi)$ & $150$ GHz \\
%  Axial frequency & $\omega_z/(2\pi)$ & $200$ MHz \\
%  Magnetron frequency & $\omega_-/(2\pi)$ & $133$ kHz \\
%  \hline
%  Cyclotron damping (free space) & $\tau_+$ & $0.09$ s \\
%  Axial damping & $\tau_z$ & $30$ ms \\
%  Magnetron damping & $\tau_-$ & $10^9$ yr \\
%  \hline
%\end{tabular}

\subsection{Attaining High Stability} \label{sec:AttainingHighStability}

Measuring the electron $g/2$ with a precision of parts in $10^{13}$ requires careful attention to making a stable trapping potential.  Even more
important is a stable magnetic field since the frequencies $\fcb$ and $\nuab$ that we measure are both proportional to $B$, and we are not able to
measure these frequencies at exactly the same time.

A major defense against external field fluctuations is a high magnetic field.  This makes fluctuations from outside sources relatively smaller. The
largest source of ambient magnetic noise is a subway that produces 50~nT (0.5 mG, 10~ppb) fluctuations in our lab and that would limit us to four
hours of data taking per day (when the subway stops running) if we did not shield the electron from them. Eddy currents in the high-conductivity
aluminum and copper cylinders of the dewars and the magnet bore shield high-frequency fluctuations~\cite{EddyCurrents}. For slower fluctuations, the
aforementioned shelf-shielding solenoid \cite{SelfShieldingSolenoid} has the correct geometry to make the central field always equal to the average
field over the solenoid cross-section.  This translates flux conservation into central-field conservation, shielding external fluctuations by more
than a factor of 150~\cite{SelfShieldingSolenoidDemonstrated}.

Stabilizing the field produced by the solenoid requires that care is taken when the field value is changed, since changing the current in the
solenoid alters the forces between windings.  Resulting stresses can take months to stabilize if the coil is not pre-stressed by ``over-currenting''
the magnet. Our recipe is to overshoot the target value by a few percent of the change, undershoot by a similar amount, and then move to the desired
field, pausing several minutes after each change.

The apparatus in Fig.~\ref{fig:Apparatus} evolved historically rather than being designed for maximum magnetic field stability in the final
configuration. Because the solenoid and the trap electrodes are suspended from widely separated support points, temperature and pressure changes can
cause the electrodes to move relative to the solenoid.  Apparatus vibrations can do the same. Insofar as the magnetic field is not perfectly
homogeneous, despite careful adjusting of the persistent currents in ten superconducting shim coils, such relative motion changes the field seen by
the electron.

To counteract this, we have long regulated the five He and N$_2$ pressures in the cryostats to $\approx50$~ppm to maintain the temperature of both
the bath and the solenoid itself~\cite{ThesisPhillips,VanDyckStability}. Recently we also relocated the dilution refrigerator vacuum pumps to an
isolated room at the end of a 12~m pipe run.  This reduced vibration by more than an order of magnitude at frequencies related to the pump motion and
reduced the noise level for the experimenters but did not obviously improve the $g/2$ data.

Because some of the structure establishing the relative location of the trap electrodes and the solenoid is at room temperature, changes in room
temperature can move the electron in the magnetic field. The lab temperature routinely cycles 1 -- 2 K daily, so we house the apparatus in a large,
insulated enclosure within which we actively regulate the air temperature to 0.1 K. A refrigerated circulating bath (ThermoNeslab RTE-17) pumps water
into the regulated zone and through an automobile transmission fluid radiator, heating and cooling the water to maintain constant air temperature.
Fans couple the water and air temperatures and keep a uniform air temperature throughout.

The choice of materials for the trap electrodes and its vacuum container is also crucial to attaining high field stability
\cite{HarvardMagneticMoment2006,ThesisOdom}. Copper trap electrodes, for example, have a nuclear paramagnetism at 0.1 mK that makes the electron see
a magnetic field that changes at an unacceptable level with very small changes in trap temperature.  We thus use only low-Curie-constant materials
such as silver, quartz, titanium, and molybdenum at the refrigerator base temperature and we regulate the mixing chamber temperature to better than
1~mK.

A stable axial frequency is also extremely important since  small changes in the measured axial frequency reveal one-quantum transitions of the
cyclotron and spin energy (as will be discussed in Sec.~\ref{sec:QNDDetection}).  A trapping potential without thermal fluctuations is provided by a
charged capacitor ($10~\mu$F) that has a very low leakage resistance at low temperature. We add to or subtract from the charge on the capacitor
using 50 ms current pulses sent to the capacitor through a $100~$M$\Omega$ resistor as needed to keep the measured axial frequency constant. Because of
the orthogonalized trap design \cite{CylindricalPenningTrap} already discussed, the potential applied to the compensation electrodes (to make the
electron see as close to a pure electrostatic quadrupole potential as possible) has little effect upon the axial frequency.

\subsection{One Electron: Its Motions and Damping}

We load a single electron using an electron beam from an atomically sharp tungsten field-emission tip. A hole in the bottom endcap electrode admits
the beam, which hits the top endcap electrode and releases gas atoms cryopumped on the surface. Collisions between the beam and gas atoms allows an
electron to fall into the trap.  Adjusting the beam energy and the time it is left on controls the number of electrons loaded.

The electron has three motions in the Penning trap formed by the $B=5.4$ T magnetic field, and the electrostatic quadrupole potential.  The cyclotron
motion in the trap has a frequency $\nucb\approx 150$ GHz.  The axial frequency, for the harmonic oscillator parallel to the
magnetic field direction, is $\nuzb\approx200$~MHz.  A circular magnetron motion, perpendicular to $\mathbf{B}$, has an oscillation frequency,
$\numb\approx133$~kHz.  The spin precession frequency, which we do not measure directly, is a part-per-thousand higher than the cyclotron frequency.  The
frequency difference is the anomaly frequency, $\nuab\approx174$~MHz, which we do measure directly.

The undamped spin motion is essentially uncoupled from its environment~\cite{Review}. The cyclotron motion is only weakly damped. By controlling the
cyclotron frequency relative to that of the cavity radiation modes, we alter the density of radiation states and inhibit the spontaneous emission of
synchrotron radiation~\cite{InhibitionLetter,Review} by 10 to 50 times the (90~ms)$^{-1}$ free-space rate. Blackbody photons that could excite from
the cyclotron ground state are eliminated because the trap cavity is cooled by the dilution refrigerator to 100~mK~\cite{QuantumCyclotron}. The axial
motion is cooled by a resonant circuit at a rate $\gz\approx(0.2~\textrm{s})^{-1}$ to as low as 230~mK (from 5~K) when the detection amplifier is
off. The magnetron radius is minimized with axial sideband cooling~\cite{Review}.

\newcommand{\QuantumJumpsFigure}{
\begin{figure}[htbp!]
\centering
\includegraphics[width=\columnwidth]{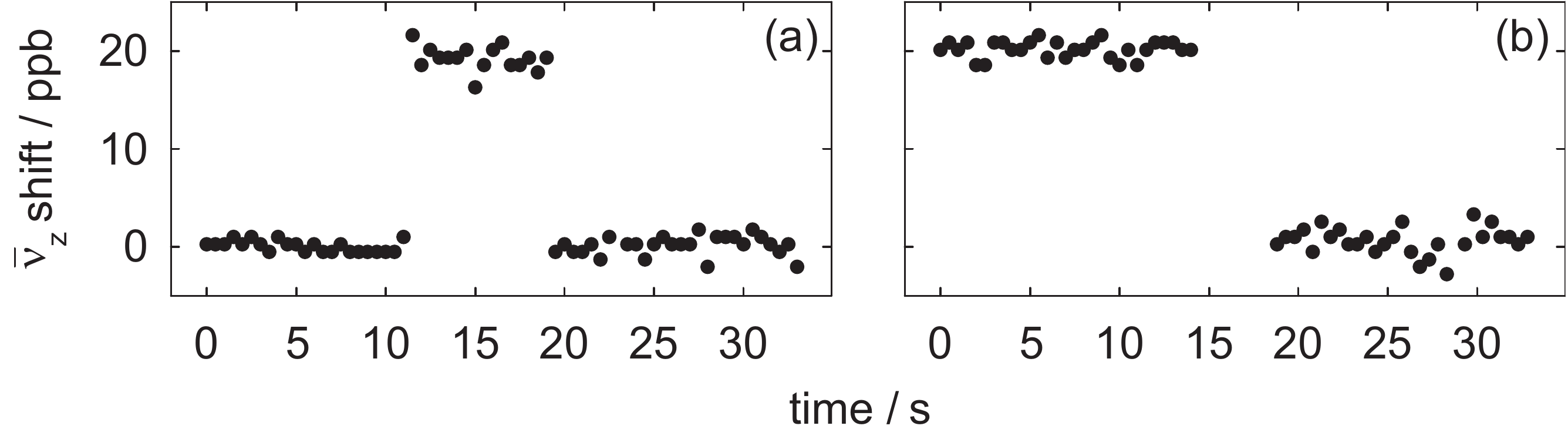}
\caption{Two quantum leaps: A cyclotron jump (a) and spin flip (b) measured in a QND manner through shifts in the axial
frequency.}\label{fig:QuantumJumps}
\end{figure}
}

\subsection{QND Detection}
\label{sec:QNDDetection}

Quantum nondemolition (QND) detection has the property that repeated measurements of the energy eigenstate of the quantum system will not change the
state \cite{QndReview}.  This is crucial for our detection of one-quantum transitions in the cyclotron motion where we do
not want the detection system to be producing the transitions that we observe.  In this section we discuss the QND coupling and in the next our
readout system.

Detecting a single 150 GHz photon from the decay of one cyclotron energy level to the level below is very difficult -- because the frequency is so
high and because it is difficult to cover the solid angle into which the photon could be emitted. Instead we get the single-photon sensitivity by
coupling the cyclotron motion to the orthogonal axial motion at 200 MHz, a frequency at which we are able to make sensitive detection electronics
\cite{ThesisDurso}. The QND nature of the detection means the thermally driven axial motion of the electron does not change the state of the cyclotron motion.

We use a magnetic bottle gradient that is familiar from plasma physics and from earlier electron measurements
\cite{DehmeltMagneticBottle,DehmeltMagneticMoment},
\begin{equation}
    \mathbf{\Delta B} = B_2\left[\left(z^2-\rho^2/2\right)\zhat - z\rho\rhohat\right],
    \label{eq:bottle}
\end{equation}
with $B_2 = 1540$~T/m$^{2}$. The gradient arises from a pair of thin nickel rings (Fig.~\ref{fig:CylindricalTrap}) that are completely saturated in the strong
field from the superconducting solenoid. To lowest order the rings modify $B$ by $\approx -0.7\%$ -- merely changing the magnetic field that the
electron experiences without affecting our measurement.

The formal requirement for a QND measurement is that the Hamiltonian of the quantum system (i.e. the cyclotron Hamiltonian) and the Hamiltonian
describing the interaction of the quantum system and the classical measurement system must commute. The Hamiltonian that couples the quantum
cyclotron and spin motions to the axial motion does so.  It has the form $-\mu B$, where $\mu$ is the magnetic moment associated with the cyclotron
motion or the spin.  The coupling Hamiltonian thus has a term that goes as $\mu z^2$. This term has the same spatial symmetry as does the axial
Hamiltonian, $H = \frac{1}{2} m (2\pi\nuzb)^2 z^2$. A change in the magnetic moment that takes place from a one-quantum change in the cyclotron or
spin magnetic moment thus changes the observed axial frequency of the suspended electron.

\QuantumJumpsFigure

The result is that the frequency of the axial motion $\nuzb$ shifts by
\begin{equation}
\Delta \nuzb = \delta_B (n + m_s),
\end{equation}
in proportion to the cyclotron quantum number $n$ and the spin quantum number $m_s$.  Fig.~\ref{fig:QuantumJumps} shows the $\Delta \nuzb = 4$ Hz
shift in the 200 MHz axial frequency that takes place for one-quantum changes in cyclotron (Fig.~\ref{fig:QuantumJumps}a) and spin energy
(Fig.~\ref{fig:QuantumJumps}b). The $20$ ppb shift is easy to observed with an averaging time of only 0.5 s.  We typically measure with an averaging
time that is half this value.

\subsection{One-Electron Self-Excited Oscillator}

Cyclotron excitations and spin flips are generally induced while the detection system is off, as will be discussed.  After an attempt to excite the
cyclotron motion or to flip the spin has been made, the detection system is then turned on to detect the state of the system. Spontaneous emission of
synchrotron radiation from the cyclotron motion is inhibited (Sec.~\ref{sec:InhibitedEmission}) to give the time that is needed.  The small axial
frequency shifts which signal changes in cyclotron quantum number or spin are measured at the required precision using a one-electron self-excited
oscillator \cite{SelfExcitedOscillator} that turns on rapidly.

The $200$ MHz axial frequency lies in the radio-frequency (rf) range which is more experimentally accessible than the microwave range of the 150 GHz
cyclotron and spin frequencies, as mentioned. Nevertheless, standard rf techniques must be carefully tailored for our low-noise, cryogenic
experiment. The electron axial oscillation induces image currents in the trap electrodes that are proportional to the axial velocity of the electron
\cite{ElectronCalorimeter,Review}. An inductor is placed in parallel with the capacitance between two trap electrodes to cancel the reactance of the
capacitor which would otherwise short out the induced signal.  The rf loss in the tuned circuit that is formed is an effective resistance that damps
the axial motion.

The voltage that the electron motion induces across this effective resistance is amplified with two cryogenic amplifiers. The heart of each amplifier
is a single-gate high electron mobility transistor (Fujitsu FHX13LG).

The first amplifier is at the 100 mK dilution refrigerator base temperature. Operating this amplifier without crashing the dilution refrigerator
requires operating with a power dissipation in the FET that is three orders of magnitude below the transistor's 10 mW design dissipation.  The
effective axial temperature for the electron while current is flowing through the FET is about 5 K, well above the ambient temperature. Very careful
heat sinking makes it possible for the effective axial temperature of the electron to cool to below 350 mK in approximately one second, taking the electron
axial motion to this temperature. Cyclotron excitations and spin flips are induced only when the axial motion is so cooled, since the electron is
then making the smallest possible excursion in the magnetic field gradient.

The second amplifier is mounted on the nominally 600 mK still of the dilution refrigerator.  This amplifier counteracts the attenuation of
thermally-isolating but lossy stainless steel transmission line that caries the amplified signal out of the refrigerator. The second amplifier boosts
the signal above the noise floor of the first room-temperature amplifier.

We feed this signal back to trap electrodes as a drive. Because the induced signal is proportional to the electron's axial velocity, this feedback alters the axial damping
force, a force that is also proportional to the electron velocity.  Changing the feedback gain thus changes the damping rate.  As the gain increases,
the damping rate decreases as does the effective axial temperature of the electron, in accord with the fluctuation dissipation theorem
\cite{FluctuationDissipationTheorem}.  The invariant ratio of the separately measured damping rate and the effective temperature has been
demonstrated \cite{FeedbackCoolingPRL}, thereby also demonstrating that the amplifier adds very little noise to the feedback.

Setting the feedback gain to make the drive exactly cancel the damping in the attached circuit could sustain a large axial oscillation
amplitude, in principle.  However, the gain cannot be perfectly adjusted and noise fluctuations will always drive the axial oscillation exponentially away from equilibrium.  We thus stabilize the oscillation amplitude using a digital signal processor (DSP) that Fourier transforms the signal,
and adjusts the feedback gain in real time to keep the signal at a fixed value.

This one-particle self-excited oscillator is turned on after an attempt has been made to excite the cyclotron energy up one level, or to flip the
spin.  The frequency of the axial oscillation that rapidly stabilizes at a large and easily detected amplitude is then measured.  Small shifts in
this frequency reveal whether the cyclotron motion has been excited or whether the spin has flipped, as illustrated in Fig.~\ref{fig:QuantumJumps}.

\newcommand{\InhibitedEmissionFigure}{
\begin{figure}[htbp!]
\centering
\includegraphics*[width=\columnwidth]{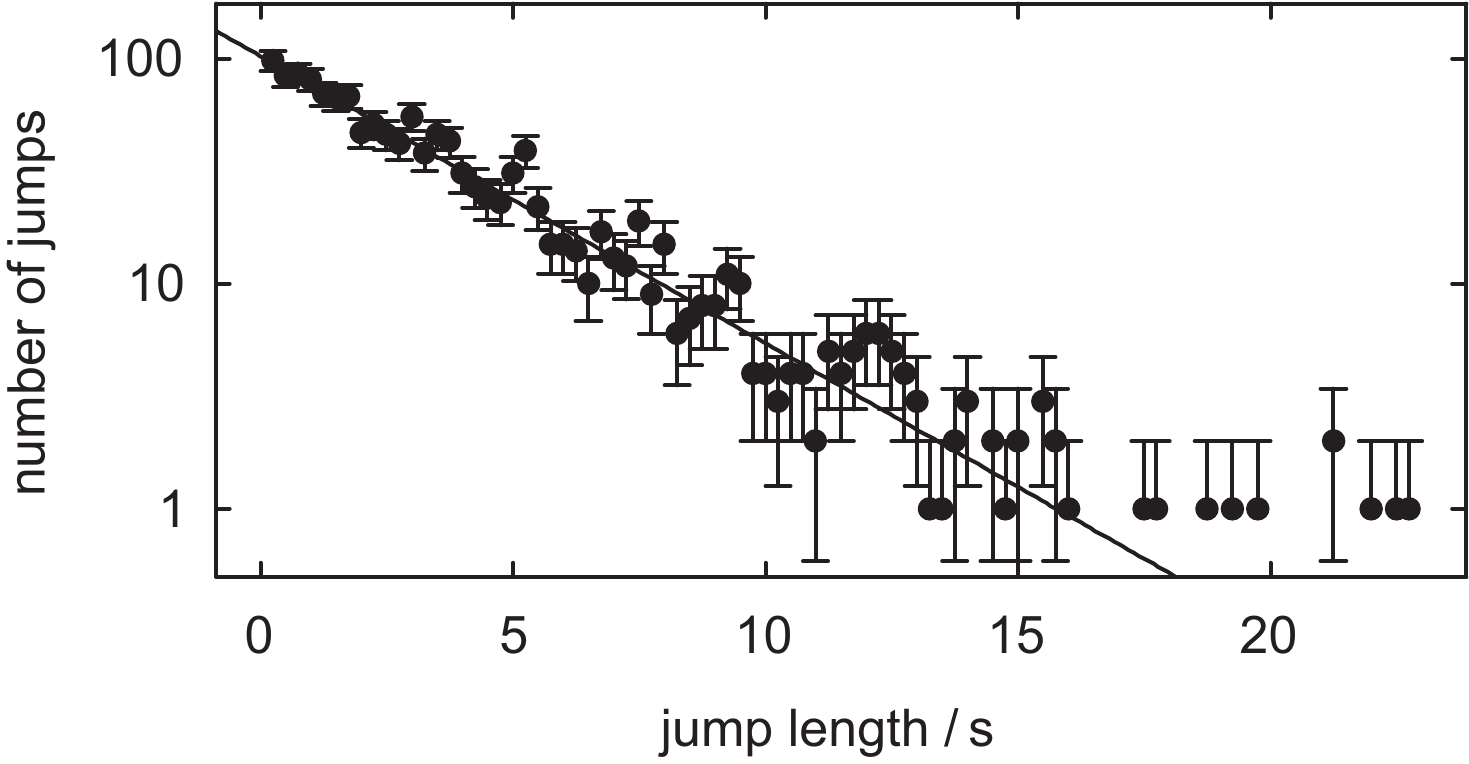}
\caption{Histogram of the time that the electron spends in the first excited state with an exponential fit. The decay time, 3.4(1)~s in this example,
depends on how close the cyclotron frequency is to neighboring radiation modes of the trap cavity.  Lifetimes as long as 16 s have been observed.
}\label{fig:InhibitedEmission}
\end{figure}
}

\subsection{Inhibited Spontaneous Emission}
\label{sec:InhibitedEmission}

One of the early papers in what has come to be know as cavity electrodynamics was an observation of inhibited
spontaneous emission within a Penning trap \cite{InhibitionLetter} -- the first time that inhibited spontaneous
emission was observed within a cavity and with only one particle -- as anticipated earlier
\cite{PurcellCavity,Kleppner1981}. As already mentioned, the cylindrical Penning trap \cite{CylindricalPenningTrap} was
invented to provide boundary conditions that would allow the control of the electron--cavity coupling, using an
understandable geometry that allows the calculation of cavity shifts to the electron's cyclotron frequency.

The spontaneous emission rate can be easily measured directly, by making a histogram of the time the electron spends in
the first excited state after being excited by a microwave drive injected into the trap cavity. Fig.~\ref{fig:InhibitedEmission} shows a sample histogram which fits well to an exponential (solid curve) with a
lifetime of 3.4 s in this example.

\InhibitedEmissionFigure

Stimulated emission is avoided by making these observations only when the cavity is at low temperature so that
effectively no black body photons are present.  The detector makes thermal fluctuations of the axial oscillation
amplitude, and these in turn make the cyclotron frequency fluctuate. For measuring the cyclotron decay time, however,
this does not matter because the fluctuations in axial amplitude are small compared to the 2 mm wavelength of the
radiation that excites the cyclotron motion.

The spontaneous emission rate into free space is \cite{Review}
\begin{equation}
\gc = \frac{1}{4\pi\epsilon_0}\frac{4 e^2}{3 m c^3}\frac{\wcb^3}{\wcb-\wmb} \approx \frac{1}{89~\rm{ms}}
\end{equation}
The measured rate in the example of Fig.~\ref{fig:InhibitedEmission} is thus suppressed by a factor of 38.  The density of states within
the cylindrical trap cavity is not that of free space.  Instead the density of states for the radiation is peaked at
the resonance frequencies of the radiation modes of the cavity, and falls to very low values between the radiation
modes. We attain the inhibited spontaneous emission by tuning the magnetic field so that the cyclotron frequency is far from radiation modes.  With the right choice of magnetic field we have increased the lifetime to 16 s,
which is a cavity suppression of spontaneous emission by a factor of 180.

In Sec.~\ref{sec:SingleElectronModeDetection} we report on using the direct measurements of the radiation rate for electron cyclotron motion
to probe the radiation modes of the cavity, with the radiation rate increasing sharply at frequencies that approach a
resonant mode of the cavity.

\subsection{Quantum Jump Spectroscopy}\label{sec:QuantumJumpSpectroscopy}

We probe the cyclotron and anomaly resonances with quantum jump spectroscopy, in which we apply a drive in discrete frequency steps, checking between
applications for a one-quantum transition and building a histogram of the ratio of excitations to attempts at each frequency.
Fig.~\ref{fig:ObservedLineshapes} show the observed lineshapes upon which our best measurement is based. We discuss how the measured points
are obtained first, and then discuss theoretical lineshapes in Sec.~\ref{sec:Lineshape}.

A typical data run consists of alternating scans of the cyclotron and anomaly lines and occurs at night, with daytime runs only possible on Sundays
and holidays when the ambient magnetic field noise is lower. Interleaved every three hours among these scans are periods of magnetic field monitoring
to track long-term drifts using the electron itself as the magnetometer. In addition, we continuously monitor over fifty environmental parameters
such as refrigerator temperatures, cryogen pressures and flows, and the ambient magnetic field in the lab so that we may screen data for abnormal
conditions and troubleshoot problems.

Cyclotron transitions are driven by injecting microwaves into the cavity. The microwaves originate as a 15~GHz drive from a signal generator (Agilent
E8251A) whose low-phase-noise, 10~MHz oven-controlled crystal oscillator serves as the timebase for all frequencies in the experiment. After passing
through a waveguide that removes all subharmonics, the signal enters a microwave circuit that includes an impact ionization avalanche transit-time
(IMPATT) diode, which multiplies the frequency by ten and outputs the $\fcb$ drive at a power of 2~mW. Voltage-controlled attenuators reduce the
strength of the drive, which is broadcast from a room temperature horn through teflon lenses to a horn at 100~mK (Fig.~\ref{fig:Apparatus}) and
enters the trap cavity through an inlet waveguide (Fig.~\ref{fig:CylindricalTrap}).

Anomaly transitions are driven by potentials, oscillating near $\nuab$, applied to electrodes to drive off-resonant axial motion through the magnetic
bottle gradient (Eq.~\ref{eq:bottle}). The gradient's $z\rho\rhohat$ term mixes the driven oscillation of $z$ at $\nuab$ with that of $\rho$ at
$\fcb$ to produce an oscillating magnetic field perpendicular to $\mathbf{B}$ as needed to flip the spin. The axial amplitude required to produce the
desired transition probability is too small to affect the lineshape (Sec.~\ref{sec:PowerShifts}); nevertheless, we apply a detuned drive of the same
strength during cyclotron attempts so the electron samples the same magnetic gradient.

Quantum jump spectroscopy of each resonance follows the same procedure. With the electron prepared in the spin-up ground state $\ket{0,\frac{1}{2}}$,
the magnetron radius is reduced with 1.5~s of strong sideband cooling at $\nuzb+\numb$ with the SEO turned off immediately and the detection
amplifiers turned off after 0.5~s. After an additional 1~s to allow the axial motion to thermalize with the tuned circuit, we apply a 2~s pulse of
either a cyclotron drive near $\fcb$ or an anomaly drive near $\nuab$ with the other drive applied simultaneously but detuned far from resonance. The
detection electronics and SEO are turned back on; after waiting 1~s to build a steady-state axial amplitude, we measure $\nuzb$ and look for a 20~ppb
shift up (from a cyclotron transition) or down (from an anomaly transition followed by a spontaneous decay to $\ket{0,-\frac{1}{2}}$) in frequency.
Cavity-inhibited spontaneous emission provides the time needed to observe cyclotron transitions before decay. The several-cyclotron-lifetimes wait
for a spontaneous decay after an anomaly attempt is the rate-limiting step in the spectroscopy. After a successful anomaly transition and decay,
simultaneous cyclotron and anomaly drives pump the electron back to $\ket{0,\frac{1}{2}}$. All timing is done in hardware with a pulse generator. We
probe each resonance line with discrete excitation attempts spaced in frequency by approximately 10\% of the linewidth. We step through each drive
frequency on the $\fcb$ line, then each on the $\nuab$ line, and repeat.

\subsection{The Electron as Magnetometer} \label{sec:EdgeTracking}

Slow drifts of the magnetic field are corrected using the electron itself as a magnetometer. Accounting for these drifts allows the combination of
data taken over many days, giving a lineshape signal-to-noise that allows the systematic investigation of lineshape uncertainties at each field. For
a half-hour at the beginning and end of a run and again every three hours throughout, we alter our cyclotron spectroscopy routine by applying a
stronger drive at a frequency below $\fcb$. Using the same timing as above but a ten-times-finer frequency step, we increase the drive frequency
until observing a successful transition. We then jump back 60 steps and begin again.
Fitting a polynomial to such cyclotron-edge tracking data allows the normalization of the raw cyclotron and anomaly data to a common magnetic field.

\subsection{Measuring the axial frequency}

Our expression for $g/2$ (Eq.~\ref{eq:THEgEQ}) requires a measurement of the axial frequency $\nuzb$. For a relative uncertainty in $g$ below 0.1~ppt, we must know $\nuzb$ to better than 50~ppb, or 10~Hz. This is easily done.

While we routinely measure $\nuzb$ during QND detection of cyclotron and spin states, this self-excited oscillation frequency includes an amplitude-dependent anharmonic shift from the low-amplitude, thermally excited axial motion during cyclotron and anomaly excitation attempts. The shift is typically a few hertz. We account for this shift by directly measuring the thermal axial frequency with the amplifiers on; the axial resonance appears as a narrow dip where the electron has ``shorted-out'' the amplifier noise~\cite{ElectronCalorimeter}. This dip frequency is negligibly different than the slightly lower-amplitude one that pertains during cyclotron and anomaly excitation, which occurs with the amplifiers off. All additional shifts from interaction with the amplifier or anomaly-drive-induced power shifts~\cite{Palmer} are negligible at our precision. Our result for $g/2$ has no uncertainty arising from the measurement of $\nuzb$.

\section{Cyclotron and Anomaly Lineshapes} \label{sec:Lineshape}

Quantum jump spectroscopy resolves the cyclotron and anomaly resonance lines, and we rely on an invariant property of the expected lineshapes (their
mean frequency) to determine $\fcb$ and $\nuab$ with uncertainties less than the linewidths. The observed lines are slightly broader than expected,
an effect we attribute to magnetic field fluctuations. Our understanding of this broadening is the primary limitation of the most accurate of our measurements.

\newcommand{\LineshapeExamplesFigure}{
\begin{figure}[htbp!]
\centering
\includegraphics*[width=\columnwidth]{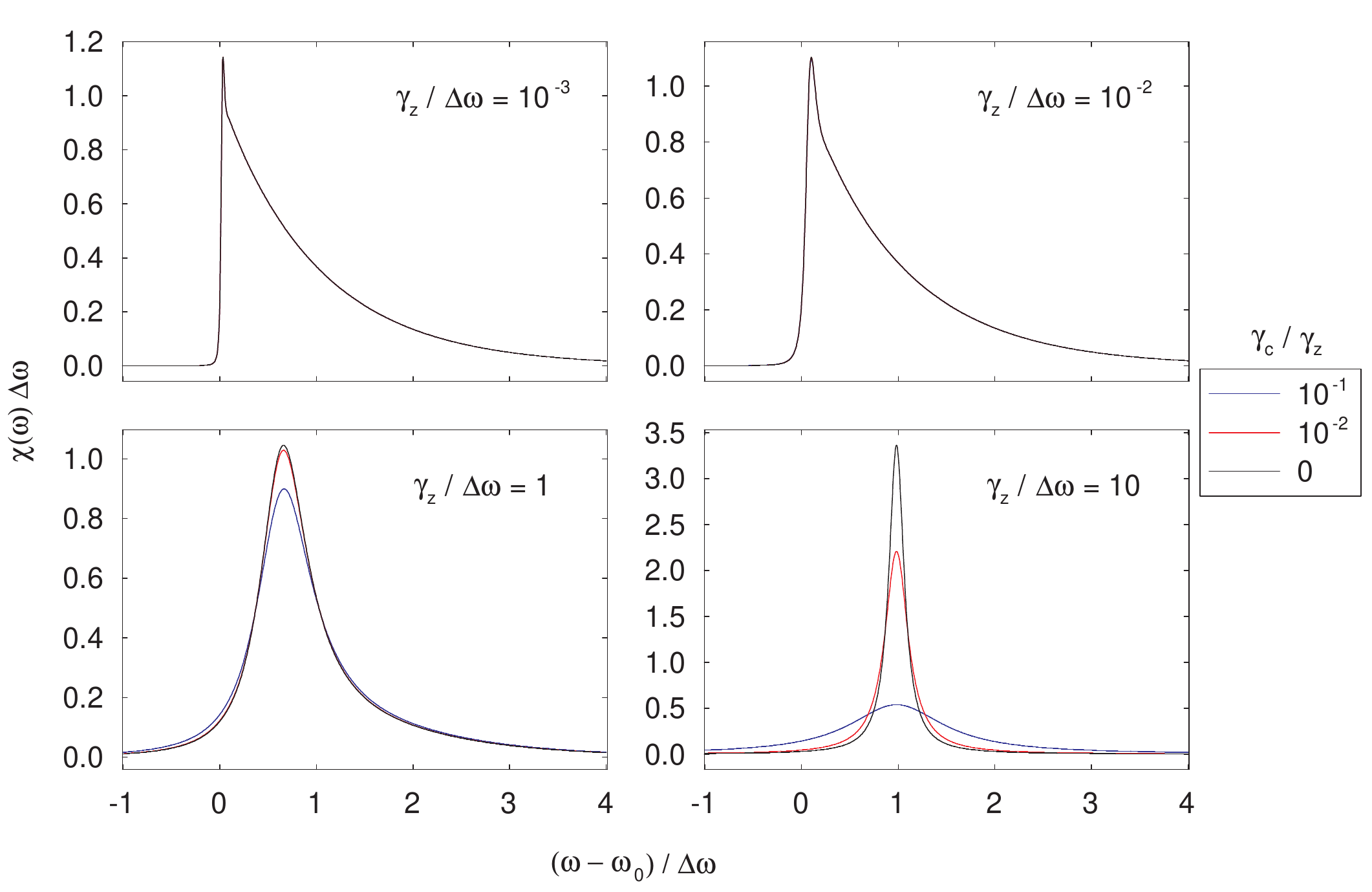}
\caption{The theoretical Brownian-motion lineshape for various $\gz/\Dw$ and $\gc/\gz$. Our cyclotron line has $\gz/\Dw \approx 10^{-2}$. Our anomaly line has $\gz/\Dw \approx 10$. For our $\gz$, $\gc/\gz = 10^{-1}$--$10^{-2}$ correspond to lifetimes of 1.6 s -- 16 s. The infinite-lifetime limit is $\gc/\gz=0$.
}\label{fig:LineshapeExamples}
\end{figure}
}

\subsection{Lineshape model}

Detailed discussion of the cyclotron and anomaly resonance lineshapes are available \cite{BrownLineshape,Review,ThesisDurso}.  Here we give a summary, stressing what must be assumed and calculated to enable the measurement of the electron magnetic moment and the uncertainties in the measurement.

For cyclotron and anomaly drives left on for a time much longer than the inverse-linewidth and inverse-axial-damping rate $\gz^{-1}$, the probability
$P$ for a transition to occur after a time $T$ is~\cite{BrownLineshape}
\begin{equation}
    P = \frac{1}{2}\left\{1-\exp\left[-\pi T \Omega^2 \chi(\omega)\right]\right\},
    \label{eq:saturatedprob}
\end{equation}
which depends on the Rabi frequency $\Omega$ and the lineshape function $\chi(\omega)$ and saturates at 1/2 for strong drives. The radiative decay of cyclotron excitations reduces the cyclotron lineshape saturation value. For a delay $t$ between the end of the drive and the beginning of the state measurement, the lineshape saturates at $\exp(-\gc t)/2$.

In general, the lineshape $\chi(\omega)$ is the Fourier transform of a correlation function $\tilde{\chi}(t)$, which is related to the statistical
average of any fluctuations in the magnetic field, $\omega(t)$ \cite{BrownLineshape}:
\begin{equation}
    \tilde{\chi}(t) = \left<\exp\left[-i \int_0^t dt^\prime \omega(t^\prime)\right]\right>.
    \label{eq:lineshapecorrelationfunction}
\end{equation}

\subsubsection{During quantum jump spectroscopy -- Brownian axial motion}
The same magnetic bottle that makes a QND coupling of the cyclotron and spin energies to the axial frequency couples the axial energy to the
cyclotron and anomaly frequencies and is the primary source of the observed lineshape. For an electron on axis ($\rho=0$), it adds a $z^2$ dependence
to the magnetic field and thus to the cyclotron and anomaly frequencies, here collectively $\omega$:
\begin{equation}
    \omega(z) = \omega_0\left(1+\frac{B_2}{B}z^2\right).
    \label{eq:lineshapebottlecoupling}
\end{equation}

When probing the cyclotron or anomaly resonance for quantum jump spectroscopy, the axial motion is in thermal equilibrium with the detection amplifier. Thus, the axial position $z$ undergoes Brownian motion. The lineshape
$\chi(\omega)$ is a statistical average of these Brownian fluctuations and takes various forms depending on the relative lengths of the
fluctuation timescale---the inverse-axial-damping time $\gz^{-1}$---and the inverse-linewidth coherence time. This linewidth roughly corresponds to the
frequency shift at the root-mean-square thermal axial amplitude:
\begin{equation}
    \Dw = \omega_0 \frac{B_2}{B} z^2_\textrm{rms} = \omega_0 \frac{B_2}{B}\frac{k T_z}{m \omega_z^2}.
    \label{eq:Dwdefinition}
\end{equation}

\LineshapeExamplesFigure

The relevant correlation function $\tilde{\chi}(t)$ is found by inserting Eq.~\ref{eq:lineshapebottlecoupling}, whose time-dependence is in the axial Brownian motion, into Eq.~\ref{eq:lineshapecorrelationfunction} to get
\begin{equation}
    \tilde{\chi}(t) = e^{-i\omega_0 t} \left<\exp\left[-i\omega_0\frac{B_2}{B}\int_0^t dt^\prime z(t^\prime)^2\right]\right>.
    \label{eq:Browniancorrelationfunction}
\end{equation}
Taking the statistical average gives three equivalent solutions for the lineshape \cite{BrownLineshape},
\begin{subequations} \label{eq:lineshape}
\begin{align}
    \chi(\omega)
            &= \frac{4}{\pi}\text{Re}\bigg[\gp\gz  \notag \\
            & ~~\times \left. \int_0^\infty dt
                    \frac{e^{i(\omega-\omega_0)t}e^{-\frac{1}{2}(\gp-\gz)t}e^{-\frac{1}{2}\gc t}}
                            {(\gp+\gz)^2-(\gp-\gz)^2 e^{-\gp t}}\right]     \label{eq:lineshape:integral}   \\
            &= \frac{4}{\pi}\text{Re}\bigg[\frac{\gp\gz}{(\gp+\gz)^2} \notag \\
            & ~~\times \left. \sum_{n=0}^\infty \frac{(\gp-\gz)^{2n} (\gp+\gz)^{-2n}}
                    {(n+\frac{1}{2})\gp+\frac{1}{2}(\gamma_c-\gz)-i(\omega-\omega_0)} \right]           \\
            &= -\frac{4}{\pi}\text{Re}\!\left[\frac{\gz}{K (\gz+\gp)^2} \right. \notag \\
            & ~~\times \left. _2F_1\!\left(1,-K;1-K;\dfrac{(\gz-\gp)^2}{(\gz+\gp)^2}\right)\right] ,
\end{align}
\end{subequations}
where
\begin{gather}
    \gp = \sqrt{\gz^2+4i\gz\Dw}, \\
    K = \frac{2i(\omega-\omega_0)+\gz-\gp-\gc}{2\gp},
\end{gather}
``Re'' denotes the real part, and $_2F_1(a,b;c;z)$ is a hypergeometric function. Examples of the lineshape for various values of $\gz/\Dw$ and $\gz/\gc$ are shown in Fig.~\ref{fig:LineshapeExamples}.

In contrast to some previous presentations of the lineshape~\cite{Review,ThesisDurso}, we have not taken the limit of low cyclotron damping.\footnote{Brown does keep a non-zero cyclotron damping rate when deriving the anomaly line, with a result~\cite[Eq. 6.12]{BrownLineshape} identical to that presented here. The derivation assumes an anomaly excitation technique that differs from ours (Sec.~\ref{sec:QuantumJumpSpectroscopy}), but the only effect is a redefinition of the Rabi frequency~\cite{Palmer}.} Our resolution of the anomaly line is fine enough that we must include the broadening from the finite lifetime of the $\ket{1,-\frac{1}{2}}$ state.

The cyclotron and anomaly lines are in two limits of this lineshape. For the cyclotron line, $\gz/\Dw\approx10^{-2}\ll 1$ and the axial motion is
essentially decoupled from the amplifier during the inverse-linewidth coherence time. During that time, the electron remains in a single axial state
and the lineshape is a Lorentzian with the natural linewidth, $\gc$, and centered on the frequency given by Eq.~\ref{eq:lineshapebottlecoupling} with
the rms axial amplitude of that state as $z$. We do not know which axial state that is, however, and since excitation attempts occur on timescales
longer than $\gz^{-1}$, subsequent attempts will be in different states. Thus, the composite lineshape after many attempts is the convolution of the
instantaneous lineshape (the narrow Lorentzian) and the Boltzmann distribution of axial states. That is, the lineshape is a decaying exponential with
a sharp edge at $\omega_0$ and a width of $\Dw$. The cyclotron line is close to this ``exponential'' limit and should have a sharp edge at the
zero-axial-amplitude cyclotron frequency that is useful for quick field measurements such as tracking drifts.

In contrast, the anomaly line has $\gz/\Dw\approx 10 \gg 1$, and the axial motion is strongly coupled to the amplifier. During the inverse-linewidth
coherence time, the axial amplitude relaxes to the thermal $z_\textrm{rms}$ yielding a lineshape that approaches a natural-linewidth Lorentzian
offset from $\omega_0$ by $\Dw$ through a Lorentzian with a width of $\gc + 2\Dw^2 / \gz$.

\subsubsection{During cyclotron lifetime measurements -- driven axial motion}

The axial motion is in thermal equilibrium during the resonance probes of a $g/2$ measurement, but it is necessarily driven during QND detection. Probing the lineshape with this detection drive on adds a coherent oscillation of $z$ to the Brownian motion considered above. The magnetic bottle coupling again translates this motion into a lineshape that depends on the amplitude of the axial oscillation. In the weak-coupling limit ($\gz/\Dw\ll 1$) that
corresponds to our cyclotron line, the lineshape is~\cite[Eq.~7.18]{BrownLineshape}
\begin{align}
    \chi_d(\omega) =&~I_0(\frac{2\sqrt{(\omega-\omega_0)\Delta_d\omega}}{\Dw})
        \theta(\omega-\omega_0) \label{eq:drivenlineshape} \\
        &\times \frac{1}{\Dw}\exp(-\frac{\omega-\omega_0+\Delta_d\omega}{\Dw}), \notag
\end{align}
where
\begin{equation}
    \Delta_d\omega = \omega_0 \frac{B_2}{B} \frac{A^2}{2} ,
    \label{eq:Dpw}
\end{equation}
$A$ is the driven axial amplitude, $\theta(x)$ is the Heaviside step function, and $I_0(x)$ is the order-zero modified Bessel function. We make extensive use of this driven lineshape in calibrating the axial oscillation amplitude in Sec.~\ref{sec:SingleElectronModeDetection}.

\subsubsection{Other magnetic field fluctuations}

Although the cyclotron line is in the exponential lineshape limit ($\gz/\Dw \ll 1$) and should have a sharp low-frequency edge, all our data show an
edge-width of 0.5--1~ppb (Fig.~\ref{fig:ObservedLineshapes}). We model this discrepancy as additional fluctuations in the magnetic field. (Some potential sources are
discussed at the end of this section.) Such fluctuations can be described by adding a noise term $\eta(t)$ to the Brownian motion axial fluctuations of
Eq.~\ref{eq:lineshapebottlecoupling},
\begin{equation}
    \omega(t) = \omega_0\left(1+\frac{B_2}{B}z(t)^2+\eta(t)\right),
\end{equation}
so that the lineshape is given by the Fourier transform of
\begin{equation}
    \tilde{\chi}(t) = \left< \exp\!\left[-i\int_{0}^{t} dt^\prime \omega_0
            \left(1+\frac{B_2}{B} z(t^\prime)^2 + \eta(t^\prime)\right)\right]\right>.
\end{equation}
For magnetic field noise that is not correlated with the axial fluctuations, the average factors into
\begin{align}
    \tilde{\chi}(t) =& e^{-i \omega_0 t} \left< \exp\!\left[-i \omega_0 \frac{B_2}{B}\int_{0}^{t} dt^\prime z(t^\prime)^2 \right]\right> \label{eq:magneticfieldnoisefactors}\\
    &\quad \times \left<\exp\!\left[-i \omega_0 \int_{0}^{t} dt^\prime \eta(t^\prime)\right]\right>. \notag
\end{align}
The first two factors are the Brownian-motion lineshape and the third is an additional noise broadening. Because of the Fourier transform convolution
theorem, the resulting noisy lineshape is the noise-free lineshape of Eq.~\ref{eq:lineshape} convolved with a noise function.

\subsubsection{Invariance of the mean frequency}

An important feature of the Brownian-motion lineshape, which we use in our primary line-splitting technique, is the independence of its mean from
$\gz$.
For the low drive strengths used in this measurement, we may expand
Eq.~\ref{eq:saturatedprob} to lowest order,
\begin{equation}
    P = \frac{1}{2}\pi T \Omega^2 \chi(\omega),
    \label{eq:unsaturatedprob}
\end{equation}
such that the excitation probability is linear in the lineshape function.
Then, the average frequency of the lineshape, $\left<\omega\right>=\int_{-\infty}^{\infty}\omega\,\chi(\omega)\,d\omega$, always corresponds to that
given by the thermal $z_\text{rms}$ in the magnetic bottle field (Eq.~\ref{eq:lineshapebottlecoupling}). That is~\cite[Eq.~1.29]{BrownLineshape},
\begin{equation}
    \left<\omega\right> = \omega_0 + \Dw .
    \label{eq:averagechi}
\end{equation}
This mean is easily verified in the exponential and Lorentzian limits described above, but applies for all $\gz/\Dw$.

Importantly, any additional magnetic field noise $\eta$ affects both cyclotron and anomaly lines identically (Eq.~\ref{eq:magneticfieldnoisefactors}). If the noise fluctuates symmetrically, that is, the mean frequency of the noise function is zero, then the mean frequencies of the lines are unchanged. A non-zero noise function average would shift both lines by proportionally the same amount; this shift would then cancel in the calculation of $g/2$ (Eq.~\ref{eq:THEgEQ}).

The large-drive saturation of
Eq.~\ref{eq:saturatedprob} invalidates this mean-frequency invariance, but we keep our excitation probabilities below 20\%, where we expect any
saturation-shift of the mean frequency to be smaller than the statistical uncertainty. We check this expectation when calculating a lineshape model uncertainty.

\newcommand{\ObservedLineshapesFigure}{
\begin{figure}[htbp!]
    \centering
    \includegraphics*[width=\columnwidth]{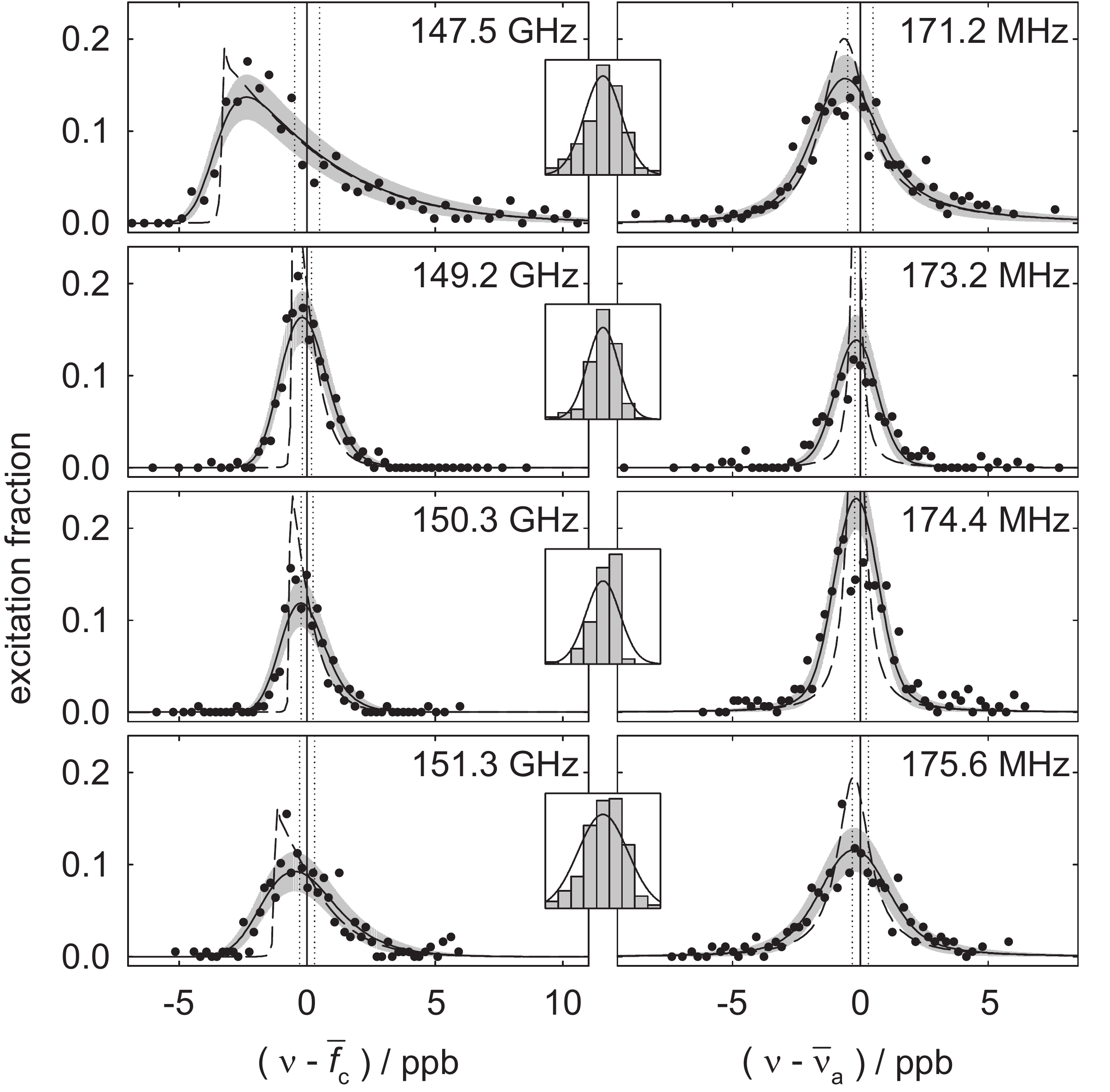}
    \caption{Quantum-jump spectroscopy lineshapes for cyclotron (left) and anomaly (right) transitions with maximum-likelihood
    fits to broadened lineshape models (solid) and inset resolution functions (solid) and edge-tracking data (histogram).
    Vertical lines show the 1-$\sigma$ uncertainties for extracted resonance frequencies. Corresponding unbroadened
    lineshapes are dashed. Gray bands indicate 1-$\sigma$ confidence limits for distributions about broadened fits.
    All plots share the same relative frequency scale.}\label{fig:ObservedLineshapes}
\end{figure}
}

\subsection{Lineshape Model Uncertainties}
\label{sec:LineshapeModelUncertainties}

We extract $\fcb$ and $\nuab$ from their resonance lines by binning the excitation attempts into histograms and calculating the weighted-mean frequency of each line
using a trapezoid-rule integration. The uncertainty in determining $\fcb$ and $\nuab$ comes from binomial uncertainties in the number of successes in
each histogram bin, and we assign it as the ``statistical'' uncertainty. Because the weighted-mean values of $\fcb$ and $\nuab$ correspond to the
same magnetic field---that at the root-mean-square thermal axial amplitude in the magnetic bottle---we may use them directly in Eq.~\ref{eq:THEgEQ}.
The assumptions in the weighted-mean method are unsaturated lines as well as identical temperature and drive conditions during cyclotron and anomaly excitations. A further assumption is a noise spectrum that either is symmetric (and thus does not shift the means) or is identical during cyclotron and anomaly excitations (and thus gives shifts that cancel in the calculation of $g/2$).

To test these assumptions, we use maximum-likelihood fits of the data to a lineshape that includes a specific model of the field-noise spectrum: the
Brownian-motion lineshape convolved with a Gaussian whose width is left as a fit parameter. These fits determine the zero-axial-amplitude $\fcb$ and
$\nuab$, which are used in Eq.~\ref{eq:THEgEQ} to calculate $g/2$. The agreement between this line-fit $g/2$ and that from the weighted mean is our
primary check on the lineshape model, and quantifying this agreement provides a systematic ``lineshape'' uncertainty.

Specifically, we compare the agreement of two weighted-mean and four fitted determinations of $g/2$. Because the weighted-mean method requires
binning the data into a histogram, we check that our result is independent of bin width by using two different numbers of bins (50 and 100) at each
field. We conduct four maximum-likelihood fits to the data: 1. a fit treating each excitation attempt separately and fitting the two lines
sequentially---cyclotron then anomaly; 2. a sequential fit to histogrammed data; 3. a simultaneous fit of both lines to histogrammed data; and 4. a
sequential fit to histogrammed data with reduced axial damping.

In three cases, we use sequential fits of the two lines because the cyclotron line better differentiates between the two main broadening mechanisms:
axial temperature and the Gaussian noise function. In the anomaly limit of the Brownian-motion lineshape, both broaden the line symmetrically; in the
cyclotron limit, the Gaussian noise function broadens symmetrically while the axial temperature broadens asymmetrically. We include one fit with a reduced
axial damping because $\gz$ is difficult to resolve precisely. Each fit contains six free parameters: the cyclotron and anomaly frequencies and Rabi
frequencies, the axial temperature, and the width of the convolved Gaussian.

\begin{table}[b]
                \caption[Summary of the lineshape model analysis]{Summary of the lineshape model analysis. All uncertainties are in ppt.}\label{tbl:LineshapeUncertainties}
    \begin{ruledtabular}
      \begin{tabular}{b{1.3 in}rrrr}
                                        \raggedleft{$\fcb$ / GHz}                                                                                        & 147.5 & 149.2 & 150.3 & 151.3 \\
                                \hline
                                        \raggedleft{\gv\ range}                                                                                                                                  & 0.73  & 0.29  & 0.33  & 0.45 \\
                                        \raggedleft{statistical uncertainty}                                                                             & 0.39  & 0.17  & 0.17  & 0.24 \\
                                                                &                                                                                                                                                                                        \multicolumn{4}{c}{$\downarrow$} \\
                                        \raggedleft{correlated lineshape model uncertainty}             & 0.24    & 0.24  & 0.24  & 0.24 \\
                                        \raggedleft{uncorrelated lineshape model uncertainty}   & 0.56         & 0             & 0.15  & 0.30 \\
                        \end{tabular}
                \end{ruledtabular}
\end{table}

At each field, we treat the 50-bin weighted-mean result as the measurement of $g/2$ and its uncertainty as the statistical uncertainty. The maximum
one-standard-deviation discrepancy, beyond this statistical uncertainty, of the other determinations is the lineshape uncertainty. To be cautious, we
avoid a reduction of the lineshape uncertainty when averaging the $g/2$ measurements at four magnetic fields by treating the minimum discrepancy
(that at 149.2~GHz) as a correlated uncertainty. It corresponds to our best understanding of the lineshape model. Any additional discrepancy is added
as an uncorrelated uncertainty. These uncertainties are summarized in Table~\ref{tbl:LineshapeUncertainties}.

\ObservedLineshapesFigure

Figure~\ref{fig:ObservedLineshapes} displays the entire dataset, representing 37 nighttime runs at four magnetic fields. The data, binned into histograms
(points), fit well to a convolution (solid curve) of a Gaussian resolution function (solid inset curve) and a Brownian-motion lineshape (dashed
curve), as indicated by one-standard-deviation confidence limits for distributions of measurements about the fits (gray bands). The vertical lines
indicate the weighted-mean frequencies and their uncertainties from both statistics and lineshape model. An additional probe of the broadening comes
from histograms (inset) of the edge-tracking data, which is used for drift normalization as described in Sec.~\ref{sec:EdgeTracking}. Although the
precise distribution of this data depends on the details of the edge-tracking procedure, simulations of our procedure indicate that it should be
distributed with a width comparable to---within a factor of two of---the Gaussian broadening width. As shown in Fig.~\ref{fig:ObservedLineshapes}, the
two agree well.

The resonance lines at 147.5~GHz and 151.3~GHz appear much broader than those at the other fields. This additional width appears to be stable
throughout the data runs at each field (otherwise the narrow lines would be much noisier at high frequencies) but varies between fields. It is
consistent with a higher axial temperature, as indicated by the asymmetric broadening of the cyclotron line with a wider exponential tail, but we
have not been able to identify the procedural differences at these fields.

These differences motivate our assignment of separate lineshape model uncertainties at each field. The agreement of the weighted-mean and fitted
$g/2$ determinations is much better for the two narrower lines than for the wider ones, which is not surprising because the wider lines rely more on
the lineshape model for line-splitting.

The weighted-mean and line-fit methods should yield a $g/2$ that is independent of axial temperature. Nevertheless, we check for any systematic
trends related to axial temperature by taking an additional set of data at 149.2~GHz with the refrigerator operating at 500~mK instead of 100~mK.
Whereas the 100~mK data fit to $T_z=0.23(3)$~K, the 500~mK data fit to 0.55(2)~K, consistent with our deliberate heating. The higher-temperature weighted-mean calculation
has a statistical uncertainty of 0.30~ppt, and the maximum-likelihood-fit checks give an uncorrelated lineshape model uncertainty of 0.46~ppt, both
larger than those of the lower-temperature data at 149.2~GHz in agreement with the temperature--uncertainty correlation noted above. Including the
statistical and uncorrelated lineshape model uncertainties, the difference between the 149.2~GHz, 500~mK $g/2$ and the 100~mK $g/2$ is 0.5(6)~ppt,
which is consistent with zero.

\subsection{Possible Broadening Sources}
\label{sec:PossibleBroadeningSources}

What could cause the additional line-broadening? We have modeled this effect above as fluctuations in the magnetic field. Attributing the line broadening to field noise assumes that the fluctuation timescale is not so fast that the noise averages away during an
excitation attempt. The relevant comparison timescale is the inverse-linewidth coherence time ($200~\mu$s for the cyclotron line and 200~ms for the
anomaly line), and any line-broadening noise must fluctuate near to or slower than these timescales. Noise-broadening from slow fluctuations is
analogous to the exponential limit ($\gz/\Dw \ll 1$) of the noise-free cyclotron line, which takes its shape from the long axial fluctuation time and
the distribution of axial energies. Our edge-tracking data provide an upper-timescale of minutes for the noise timescale because we see no
correlation between adjacent edge-tracking points, which come at intervals of several minutes. This range of allowed timescales constrains the
possible fluctuation mechanisms.

We have considered and ruled out several possible sources. Section~\ref{sec:AttainingHighStability} includes several known producers of field noise---the local subway, cryostat pressure changes, ambient temperature
changes, and temperature-dependent paramagnetism---and our efforts to reduce them. Phase noise on the cyclotron and anomaly drives would mimic field fluctuations, though the
similar broadening observed on each line is not what one would expect given the vastly different drive frequencies. Estimates based on the microwave
signal generator's specified phase noise and additional noise from an ideal multiplier suggest any frequency deviations should be over two orders of
magnitude below the level required to explain the cyclotron broadening.

Remaining candidates include relaxation of stresses in
the solenoid windings or vibration of the trap electrodes in an inhomogeneous magnetic field. Even with the magnetic field tuned to its specified
homogeneity, a 100~$\mu$m motion of the trap electrodes would cause a 0.1~ppb field variation. Such a motion could be caused by vibrations that drive
the dilution refrigerator like a 2.2-m-long pendulum.

One additional source of broadening is a distribution of magnetron radii in the $-\rho^2\zhat/2$ portion of the magnetic bottle
(Eq.~\ref{eq:bottle}). Ideally, our sideband cooling procedure will produce a thermal distribution of magnetron energies with a temperature related
to the axial temperature by~\cite{Review}
\begin{equation}
    T_m = - \frac{\numb}{\nuzb} T_z .
    \label{eq:magnetrontemperature}
\end{equation}
(The negative sign indicates that the magnetron degree of freedom is unbound.) Such an ideal distribution would broaden the cyclotron line by
$\numb/\nuzb$ times the axial broadening $\Dw$, which is $10^2$--$10^3$ times too small to account for the observed width. Nevertheless, we have
checked that the broadening is not due to a thermal distribution of magnetron energies by deliberately sideband cooling with the amplifiers on
($T_z=5$~K) before damping the axial energy to 230~mK; we found no change in the broadening. Sideband cooling with the coherent distribution of axial
states found during self-excitation would produce a coherent distribution of magnetron states with a width that could explain the broadening, but the
electron is not self-excited during sideband cooling. Interestingly, prior studies of sideband cooling could only achieve cooling to energies 400
times the ideal limit~\cite{Review}. If this non-ideal energy limit corresponds to a distribution of energies as well, it could explain the observed
broadening; though, it is unclear what mechanism limits the cooling and whether it applies to our setup as well.

Our knowledge of the lineshape model provides the largest uncertainty in recent $g/2$
measurements~\cite{HarvardMagneticMoment2006,HarvardMagneticMoment2008}. Future efforts will focus on explaining or eliminating the lineshape
broadening.

\newcommand{\lmnp}{\lambda_{mnp}}
\newcommand{\wmnp}{\omega_{mnp}}
\newcommand{\QTE}{Q_\textrm{TE}}
\newcommand{\QTM}{Q_\textrm{TM}}
%  Old (M)Q and (E)Q below
%\newcommand{\QTE}{^{(E)}\!Q}
%\newcommand{\QTM}{^{(M)}\!Q}

\section{Cavity Control} \label{sec:CavityShifts}

\subsection{Overview}

The averaging time needed to observed one-quantum cyclotron transitions is obtained by
inhibiting \cite{PurcellCavity,InhibitionLetter} the spontaneous emission of synchrotron radiation.  Inhibition by factors of
200 and more is accomplished with a cylindrical Penning trap cavity that was invented for
this purpose \cite{CylindricalPenningTrap}. The cylindrical trap shapes the radiation
field within the interior of a conducting right circular cylinder, modifying the density
of radiation states in a way that can be studied and understood, at the same time as it
provides the high quality electrostatic quadrupole potential needed to detect one trapped
electron with good signal-to-noise \cite{CylindricalPenningTrapDemonstrated}.

The extremely useful inhibition of spontaneous emission comes at a cost insofar as
coupled  oscillators -- the cyclotron oscillator and the radiation mode oscillators --
pull each other's frequency.  The challenge that thus arises is that the interaction of
the electron and the cavity modes also shifts the cyclotron frequency
\cite{RenormalizedModesPRL,HarvardMagneticMoment2006}.  Typical cavity shifts are at the
ppt-level in the cyclotron frequency and ppb-level in the anomaly frequency -- large enough to unacceptably shift the value of $g/2$.
The cylindrical trap geometry was selected to provide familiar and well understood
boundary conditions within which the properties of the radiation field, and hence the
shifts that these fields produce, could be calculated and understood.

The Penning trap establishes the boundary conditions of a right circular cylinder for
microwaves within the trap -- determining the electric and magnetic field of the
transverse electric and magnetic modes.  Of course, the boundaries are not perfect because the
electrodes are deliberately slit so that sections of the cavity can be separately biased
trap electrodes, because the electrodes contract as the trap is cooled from 300 to 0.1~K,
and the electrodes are not perfectly machined and aligned. The result is that the
resonant frequency and damping factor for each radiation mode are slightly shifted
and must be measured  within the cold trap cavity.

We used two independent methods to investigate the radiation modes of the cylindrical
trap cavity \textit{in situ}. For our 2006 measurement, the synchronization of the
collective motion of many electrons
\cite{SynchronizedElectronsPRL,SynchronizedElectronsPRA} was used to trace out the
cyclotron damping rate as a function of the electron cyclotron frequency, allowing
us to identify and label the transverse electric and transverse magnetic modes.  For the
2008 measurement, the one-electron cyclotron damping rate was directly measured as a function of both the
cyclotron frequency and the amplitude of the axial oscillation through the standing wave
field of the cavity modes.  We demonstrate a 165(4)~$\mu$m axial offset between the
electrostatic center of the trap and the center as defined by the standing wave fields,
and place a limit of $\rho < 10~\mu$m on any radial offset.

The electric and magnetic fields of these modes are used as input for an  analytic
calculation that is properly renormalized to avoid self-energy infinities
\cite{RenormalizedModesPRA,CavityShiftsQEDBook}.  The calculation is then adapted in a
semi-empirical way to better describe the way that the electron cyclotron frequency is
shifted and damped.  For the 2006 measurement we were able to reduce the cavity shift
uncertainty to 0.39~ppt \cite{HarvardMagneticMoment2006}.  For the 2008 measurement the
cavity shift uncertainty was reduced by an additional factor of six, to 0.06~ppt
\cite{HarvardMagneticMoment2008}.

\subsection{Mode detection with synchronized electrons}

\begin{figure}
    \centering
    \includegraphics[width=\columnwidth]{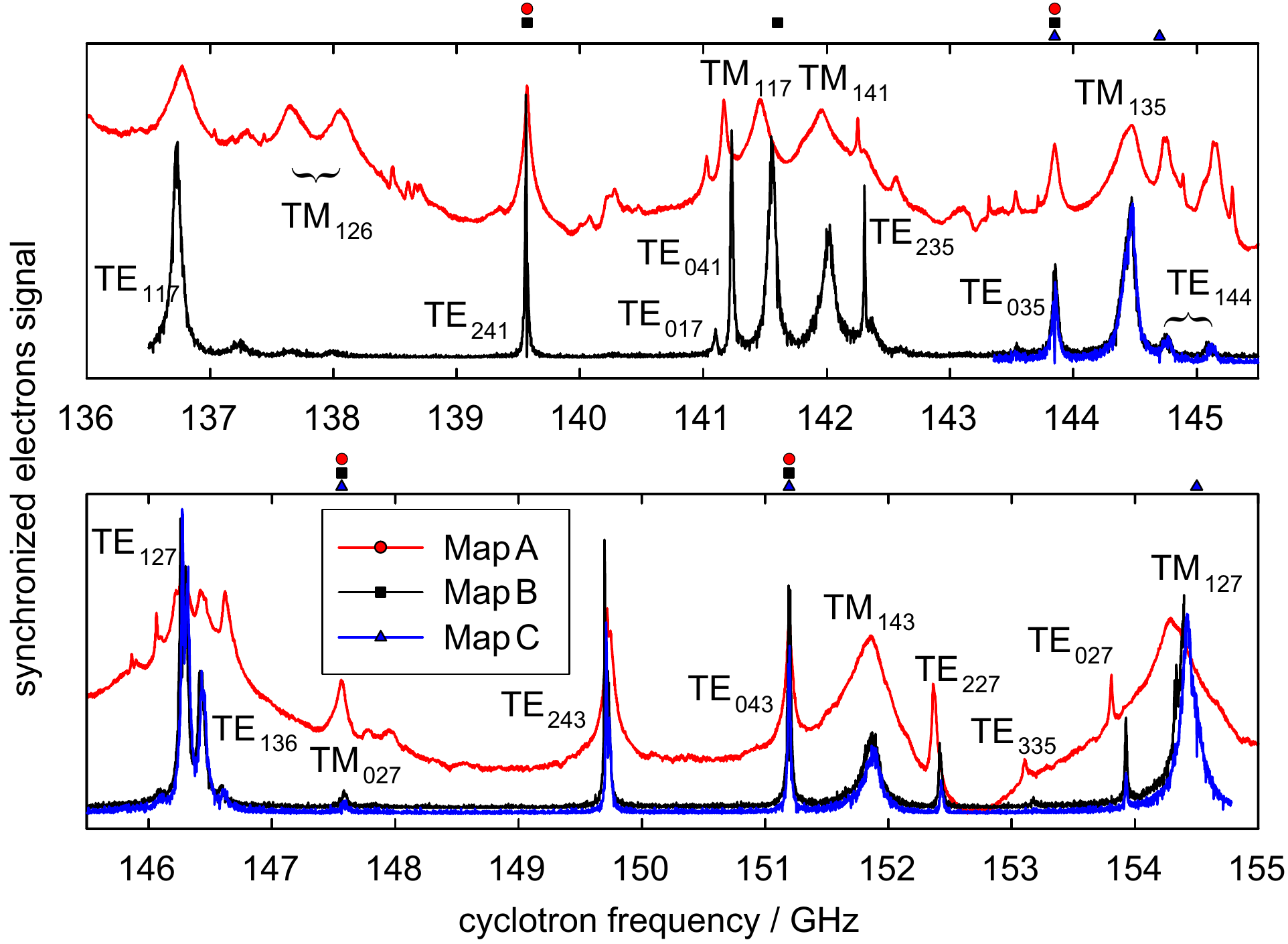}
    \caption{Modes of the trap cavity as observed with synchronized electrons in three separate sweeps.
    The calibration frequencies are indicated with symbols above the plots. The striking difference between
    Maps B and C and Map A is discussed in the text.}\label{fig:modemap}
\end{figure}

Our first technique for probing the radiation modes of the not-quite-ideal trap cavity
uses the synchronized  axial motion of approximately $2\times10^4$ electrons trapped
near its center.  When we first developed this method \cite{SynchronizedElectronsPRL}, we were able to use it
to identify more than 100 radiation mode of a cylindrical trap cavity, with frequencies
between 20 GHz and 160 GHz, with those of an ideal right circular cylinder.  The cavity
used then was nearly identical to the one used for the 2006 and 2008 measurements.  The method
seemed ideal in that it produced \textit{in situ} a spectrum with peaks at the frequency
of radiation modes (Fig.~\ref{fig:modemap}), and with lineshapes that were Lorentzian and
independent of the number of electrons used (as long as saturation and strongly coupled
regimes were avoided).  We thus interpreted these widths as the inverse of the mode
quality or $Q$ factors.

The electrons are excited by modulating the potential applied to the bottom endcap
electrode at a  frequency that is nearly twice the axial oscillation frequency of the
electrons, $2 \nuzb$. This parametric modulation of the trapping potential heats the
electrons. The axial oscillation of their center-of-mass grows exponentially but this
growth is limited to a steady-state value because the trapping potential is anharmonic
for large oscillation amplitudes. Our prior work~\cite{SynchronizedElectronsPRL,SynchronizedElectronsPRA} showed that the measured oscillation amplitude was large when the
cyclotron damping rate was high -- when the cyclotron frequency of the electrons was
resonant with a radiation mode in Fig.~\ref{fig:modemap}. The shape of the spectrum was remarkably independent of the number of electrons $N$,
which could be varied to change the cyclotron damping rate of the center-of-mass motion
$N\gc$.  In the strong coupling regime, where $N\gc$ exceeded the damping rate of the
cavity modes, the observed peaks split into two.  The axial motion of
the electrons also generates motional sidebands in the observed cavity spectra.  The
method was very robust and reproducible.

We learned a great deal about how this method worked by varying the number of electrons
used from $10^5$ electrons \cite{SynchronizedElectronsPRL,SynchronizedElectronsPRA} down
to only $2$ \cite{StochasticPhaseSwitching}, and by varying electrostatic anharmonicity,
the parametric drive strength and the slow rate at which the cyclotron frequency swept
through resonance with the radiation modes.  For electrons driven at twice their
preferred oscillation frequency, the resulting axial oscillations can have either of two
oscillation phases that differ by $\pi$.  Between resonances with the cavity radiation
modes, where the cyclotron damping was very small, the electrons oscillated in equal
numbers with both of the two axial oscillation phases, so that the detected signal from
the axial center-of-mass is very small.  On resonance with the cavity radiation modes,
with very strong cyclotron center-of-mass damping, the electrons synchronize into axial
oscillations with predominately one of the two possible phases, producing a large
detected signal.

This method is not yet fully understood despite
boundary conditions that are very carefully controlled. We have written down what we
believe are the full equations of motion for the $N$
electrons~\cite{SynchronizedElectronsPRA}, but have not obtained a solution
that represents the simple and striking behavior that is observed. One additional clue
may be that the method worked robustly over a long time in an apparatus for which
the trap's vacuum enclosure and the detection electronics were submerged in liquid
helium.  In a brief trial, we did not succeed in getting robust performance in an
apparatus in which the trap vacuum container and the detection electronics were cooled by
thermal contact to a liquid helium dewar. Perhaps the temperature and noise of the FET
detectors are more important than initially supposed.

In the apparatus used for the $g/2$ measurements, cooled to 0.1 K rather than to 4.2 K and with improved detection
electronics, the interpretation of these cavity mode spectra has become much less clear.
For causes that we have not yet discerned, we sometimes measure spectra that are like
what we measured in the first studies of the cavity spectra
\cite{SynchronizedElectronsPRA} (blue traces in Fig.~\ref{fig:modemap}). However, much
more often we measure severely broadened spectra (red traces in Fig.~\ref{fig:modemap}).

The observed cavity spectra nonetheless give us an important \textit{in situ} probe into
the spectrum of the cavity radiation modes. It
takes several hours to measure a spectrum.  In light of the striking difference
(mentioned above and discussed in more detail below) the broadened spectra marked A in
Fig.~\ref{fig:modemap} is not used in the cavity shift analysis. The 2008 measurements
uses the same trap cavity and measured spectra as in 2006
\cite{HarvardMagneticMoment2006}, but our understanding has improved due to the addition
of a new probe of the cavity spectra (next section) and through more thorough analysis.

The frequency axis for the cavity spectra is calibrated during the sweep by applying a
strong cyclotron drive at discrete frequencies (symbols in Fig.~\ref{fig:modemap}). When
resonant with $\nucb$, the magnetic-bottle coupling shifts $\nuzb$ and the parametric
drive is no longer resonant, producing a sharp dip in the signal such as that on
TE$_{241}$ near 139.6~GHz. The cyclotron frequencies for the remainder of the map are
assigned with a linear interpolation function between these discrete calibration points.
(Map A is calibrated with an earlier technique that involved stopping the magnetic field
where marked and measuring $\nucb$ directly.) Comparing the mode frequencies with those
of an ideal cylinder with dimensions similar to the trap dimensions allows us to identify
and label the TE and TM modes.

Three features aid the interpretation of the cavity
spectra~\cite{SynchronizedElectronsPRA}.
\begin{enumerate}
\item Strong cloud--mode coupling can split the Lorentzian response into a pair of normal
modes. The $Q$ of the split peaks is higher than that of the mode, making it difficult to
estimate the mode $Q$ from parametric mode maps alone.
\item The large axial motion of
the cloud during the measurement (limited only by trap anharmonicity) probes the
electric-field profiles of the modes and amplitude-modulates $\gc$, producing
axial-frequency sidebands at $\pm\nuzb$ for modes with a node at the trap center (e.g.
TE$_{144}$ at 145.0~GHz) and $\pm 2\nuzb$ for modes with an antinode at the trap center.
These motional sidebands are further explored below for the single-electron case.
\item Non-zero cloud size or a relative offset between the electrostatic center and the mode
center allows coupling to modes with nodes at the mode center (e.g. TE$_{136}$ at
146.4~GHz and TE$_{243}$ at 149.7~GHz).
\end{enumerate}

For the four magnetic fields at which we measure $g/2$, the modes with the largest
influence on the cavity shift are TE$_{127}$, TE$_{136}$, and TM$_{143}$.
Table~\ref{tbl:lifetimeparameters} summarizes the mode frequencies and $Q$ values
obtained from Lorentzian fits to the cavity spectra B and C of Fig.~\ref{fig:modemap}
after accounting for any normal-mode splitting.  The listed uncertainties are what is
estimated as part of the nonlinear least squares fitting to Lorentzian peaks, and does
not include any additional systematics contribution.

\begin{table}[h]
        \caption{Comparison of the mode parameters from the multi-electron parametric mode maps and the single-electron lifetime fits.}\label{tbl:lifetimeparameters}
        \begin{ruledtabular}
      \begin{tabular}{rc|cc}
                && parametric & lifetime  \\
                \hline
                \multirow{2}{*}{TE$_{127}$} &
            $\nu_c$\,/\,GHz & 146.289(7) & 146.322(13)  \\
                & $Q$ & 4600(900) & 4900(300)  \\ \cline{2-2}
                \multirow{2}{*}{TE$_{136}$} &
            $\nu_c$\,/\,GHz & 146.436(7) & 146.415(2)  \\
                & $Q$ & 2200(60) & 4800(200)  \\ \cline{2-2}
                \multirow{2}{*}{TM$_{143}$} &
            $\nu_c$\,/\,GHz & 151.865(4) & 151.811(16)  \\
                & $Q$ & ~890(10) & 1270(70)  \\ %\cline{2-2}
            \end{tabular}
        \end{ruledtabular}
\end{table}

In light of the differences between the cavity spectra measured when this method was
first developed, and what we now are able to observe, and because we are not currently
always able to robustly produce such spectra, we have carefully reflected upon our
measurement techniqe. Early studies on mode-detection with synchronized
electrons~\cite{SynchronizedElectronsPRA} found the CM amplitude closely followed $\gc$
for wide ranges of parameters, yielding Lorentzian mode profiles except in cases of
normal-mode splitting or motional sidebands.  With our current apparatus we find three
classes of behavior:
\begin{enumerate}
\item Convincing cavity spectra with the detected center-of-mass signal tracking $\gc$ with Lorentzian mode profiles
and no center-of-mass motion far from modes, as seen in B and C of Fig.~\ref{fig:modemap}.
\item Broadend cavity spectra in which the signal never disappears but increases and decreases with $\gc$, as seen in A.
\item No cavity spectrum at all.  With nominally the same parameters as for
the previous two cases we do not see the detected signal change as the magnetic field is
varied to sweep the cyclotron frequency. This is the most common behavior.
\end{enumerate}

There are substantial differences between the earlier apparatus (Ref.~\cite{SynchronizedElectronsPRA}) and that used for the 2006-2008 measurement, even though the trap
electrodes themselves are essentially the same.
\begin{enumerate}
\item Forty-times lower cavity temperature should increase the inter-particle Coulomb interaction
and enhance collective
motion~\cite{ElectronPlasmaLiquidCrystal,StatisticalPhysicsDensePlasmas}
\item Ten-times deeper axial potential should have a similar effect.
\item Improved detection electronics \cite{ThesisDurso} should offer greater detection
sensitivity while producing less noise to heat the trapped electrons.
\item Ten-times larger electron clouds are used because we find that smaller clouds rapidly saturate, with all
electrons in one of the two oscillation phases.
\item More heavily filtered and noise-free electrical environment should reduce noise driven transitions between the
bistable states.
\end{enumerate}

The three maps of Fig.~\ref{fig:modemap} were taken using the same trap cavity over the course of 18 months, during
which time the electrodes were thermally cycled to room temperature several times, the refrigerator was inserted and
removed, and the magnet was quenched with the electrodes inside. At no point were the electrodes themselves
disassembled or adjusted. The general alignment of the features and the precise alignment of the calibration points in
Fig.~\ref{fig:modemap} indicate that the trap cavity and its resonant modes are robust against stresses and thermal
cycles. Any misalignments in the location of a particular mode may be attributed to the calibration process,
specifically to nonlinear charging rates from the power supplies, rather than to real shifts in the mode frequencies.
This consistency suggests that the cavity itself is stable, though the variety of parametric behaviors discussed in the
previous paragraph motivates our use of an independent, one-electron mode detection technique.

\subsection{Cavity Coupling to a Single Electron}

\subsubsection{Overview}
The expressions describing the coupling of a single electron to the
electromagnetic modes lie at the heart of our cavity analysis. Their
importance is twofold: first, they are required for the calculation
of the cyclotron frequency shifts, $\Dwc$, and second, measurements
of the cyclotron damping rate, $\gc$, allow a characterization of
the cavity mode structure independent of the multi-electron
technique of the previous section. In this section, we present two
formulas for the cavity-induced cyclotron frequency shift and
damping rate. They differ in their treatment of the electron's axial
motion. Equation~\ref{eq:renormresult} applies for an electron with
negligible axial motion, corresponding to the cold thermal axial
distribution during cyclotron or anomaly excitation. Equation~\ref{eq:modeaxialamplitude} applies for a larger
axial motion that begins to probe the mode's standing wave,
corresponding to the self-excited axial state during detection and
measurement of $\gc$.

%add something to motivate the off-center part
%objectives: how to calculate, make it plausible

\subsubsection{Single-mode approximation}

Before beginning a full calculation of the cyclotron motion--cavity
coupling, it is worth modeling the interaction between the electron
and a single nearby mode, here denoted M, to give an indication of
the character of the electron--mode coupling. This approximation
will eventually aid in modeling the coupling including axial motion
in Sec.~\ref{sec:axialamplmodecoupling}. The interaction may be
approximated as that of two coupled oscillators with the resulting
electron frequency shift and damping rate given by \cite{CavityShiftsQEDBook}
\begin{subequations} \label{eq:singlemodecoupling}
\begin{align}
    \Dwc                 &= \frac{\gamma_\textrm{M}}{2}\frac{\delta}{1+\delta^2} \\
    \gc        &= \gamma_\textrm{M}\frac{1}{1+\delta^2} .
\end{align}
\end{subequations}
Here, $\gamma_\textrm{M}$ is the cyclotron damping rate when the
electron is exactly resonant with the mode and $\delta$ is the
relative detuning, defined as
\begin{equation}
    \delta=\frac{\wcb - \omega_\textrm{M}}{\Gamma_\textrm{M}/2}.
    \label{eq:modedetuning}
\end{equation}
The mode full-width at half-maximum, $\Gamma_\textrm{M}$, arises
because of losses in the cavity and may be written in terms of a
quality factor, $Q_\textrm{M}$, with the definition:
$Q_\textrm{M} = \omega_\textrm{M} / \Gamma_\textrm{M}$. The
cyclotron frequency is maximally-shifted by $\pm\gamma_\textrm{M}/4$
at $\delta=\pm1$. Furthermore, provided the cyclotron frequency is
detuned far enough from a mode that $\delta\gg1$, i.e.,
$(\wcb-\omega_\textrm{M})/\omega_\textrm{M}\gg1/(2Q_\textrm{M})$,
the shift $\Dwc$ is $Q$-independent.

It is tempting to expand on the single-mode approximation above by
adding the contributions of many modes. This mode-sum approach is
fundamentally flawed because the real part is
infinite~\cite{RenormalizedModesPRA}. A linear divergence arises
from the inclusion of the electron self-field contribution to the
cavity radiation rather than only the field reflected from the
walls. A calculation that explicitly removes the electron self-field
from the cavity standing wave, i.e., ``renormalizes'' the field,
yields a finite result and is the subject of the next section.

\subsubsection{Renormalized calculation of cyclotron--cavity coupling}

While it is possible to tackle the full cylindrical cavity directly,
removing the electron self-field from such a calculation is
difficult. It simplifies when first analyzing the interaction of the
electron with two parallel plates then adding the contribution of
interactions with the cylindrical wall. The calculation is quite involved. It is presented for a centered
particle ($z,\rho=0$) in Ref.~\onlinecite{RenormalizedModesPRA}.
We extend the calculation to any position in the cavity. This extension is required for two reasons:
our electrostatic quadrupole suspends the electron slightly
offset axially from the cavity mode center (see Sec.~\ref{sec:SingleElectronModeDetection}), and the electron's axial motion modulates the electron--cavity coupling when measuring the cyclotron damping rate (see Sec.~\ref{sec:axialamplmodecoupling}). The
derivation for arbitrary position is nearly identical to that in
Ref.~\onlinecite{RenormalizedModesPRA} except we keep the terms that
vanish for $z, \rho=0$. Rather than rehash the lengthy calculation,
we state the results and discuss some important characteristics.

For two parallel conducting
plates, i.e., the cylindrical cavity with $\rho_0 \rightarrow
\infty$, the boundary conditions may be satisfied with a series of
image charges. Renormalizing this sum is trivial---simply omit the
electron's contribution and leave that of the image charges, a
result we describe below and call $\Sigma_P$. Proceeding to the
calculation of the full cylindrical cavity and omitting the
contribution from the endcaps leaves only the correction from the
cylindrical wall, a result we call $\Sigma_S$. The final result is
the sum of the contributions from the endcaps and the wall.

At a cavity-shifted
cyclotron frequency $\omega$, the frequency shift $\Dwc = \omega - \wcb$ and damping
rate $\gc$ is given in terms of these two contributions ($\Sigma_P$, $\Sigma_S$), the free
space damping rate $\gamma_{c0}$, and a quality factor $Q$ for all
modes:
\begin{align}
    \Dwc &- \frac{i}{2}\gc = -\frac{i}{2}\gamma_{c0}    \label{eq:renormresult} \\
        &+ \omega\left\{\Sigma_S\!\left[(1+\tfrac{i}{2 Q})\omega,z,\rho\right]+\Sigma_P\!\left[(1+\tfrac{i}{2 Q})\omega,z\right]\right\} \notag .
\end{align}
The correction to $g/2$ (Eq.~\ref{eq:THEgEQ}) is equal to the relative shift of the
cyclotron frequency:
\begin{equation}
        \frac{\Delta g_{cav}}{2} = \frac{\Dwc}{\omega} .
\end{equation}
We use this
formula in a slightly modified form to calculate the cavity shifts
of the cyclotron frequency (see the discussion in the next subsection).

The renormalized calculation begins by modeling the effect of the
cavity on the electron as an electric field
$\mathbf{E^\prime}(\mathbf{r})$ arising from image charges in the
walls. It modifies the transverse equation of motion to read
\begin{equation}
    \dot{\mathbf{v}}-\boldsymbol{\omega_c}\times\mathbf{v}+\frac{e}{m}\nabla V(\mathbf{r})+\frac{1}{2}\gamma_{c0}\mathbf{v}=\frac{e}{m}\mathbf{E^\prime}(\mathbf{r}) .
    \label{eq:radialeom}
\end{equation}
The longitudinal part of $\mathbf{E^\prime}(\mathbf{r})$ gives a
negligible correction to the trapping potential
$V(\mathbf{r})$~\cite{RenormalizedModesPRA}, but the transverse part
generates the anticipated effects. Using the radiation gauge,
$\nabla\cdot\mathbf{A}=0$, the electric field may be written as the
time derivative of the vector potential. This vector potential
satisfies the wave equation with a transverse current source and
thus may be written as the convolution of that source---the moving
electron---and a Green's function subject to the appropriate
boundary conditions, see e.g.,~\cite[Sec.\,6.3-6.4]{Jackson3rdEd}.
Combining the two transverse velocity components as $v=v_x-i v_y=v_0
e^{-i\omega t}$, one can then write Eq.~\ref{eq:radialeom} as
Eq.~\ref{eq:renormresult} where $\Sigma_P$ and $\Sigma_S$ are
proportional to Fourier transforms of the part of the Green's
function that arises due to the presence of the cavity walls. Note
that $\Sigma_P$ and $\Sigma_S$ are in general complex, with the real
portion corresponding to a frequency shift and the imaginary portion
to a modified damping rate.
%\begin{equation}
%   \Dwc - \frac{i}{2} \gamma + \frac{i}{2} \gamma_c = -\omega r_0 \tilde{D}^\prime_{xx}(\omega;\mathbf{r},\mathbf{r}) ,
%   \label{eq:radialeomgreen}
%\end{equation}
%where $r_0 = e^2/(4\pi\epsilon_0 m c^2)$ is the classical electron radius. $\tilde{D}^\prime_{kl}(\omega;\mathbf{r},\mathbf{r^\prime})$ is the Fourier transform of the part of that Green's function that arises due to the presence of the cavity walls. That is, it explicitly excludes the electron self-field. Note that $\tilde{D}^\prime_{xx}(\omega;\mathbf{r},\mathbf{r})$ is in general complex, with the real portion corresponding to frequency shifts and the imaginary portion to a modified damping rate.

As mentioned above, the method of images gives the parallel-plate
contribution to the cyclotron frequency shift and damping rate. As a
function of axial position $z$ and cyclotron frequency $\omega$, it
is
% Form that displays the equation as two lines in one column
%\begin{gather}
%   \Sigma_P(\omega,z) = -r_0\left[2 \sum_{j=1}^\infty F(4 j z_0) \right.\label{eq:SigmaP} \\
%                                               \left. -\sum_{j=1}^\infty F(2(2j-1)z_0 +2z)-\sum_{j=1}^\infty F(2(2j-1)z_0 - 2z)\right] . \notag
%\end{gather}
%
% Form that displays the equation across both columns
\begin{widetext}
\begin{equation}
    \Sigma_P(\omega,z) = -r_0\left[2 \sum_{j=1}^\infty F(4 j z_0) -\sum_{j=1}^\infty F(2(2j-1)z_0 +2z)-\sum_{j=1}^\infty F(2(2j-1)z_0 - 2z)\right] , \label{eq:SigmaP}
\end{equation}
\end{widetext}
where $r_0 = e^2/(4\pi\epsilon_0 m c^2)$ is the classical
electron radius and $F(z)$ is the Fourier transform of the
aforementioned Green's function a distance $\abs{z}$ from an
electron or image charge:
\begin{equation}
F(z) = \frac{1}{\abs{z}}\left[e^{i\omega\abs{z}/c}
                                            \left(1+\frac{i c}{\omega\abs{z}}-\frac{c^2}{\omega^2 z^2}\right)
                                            +\frac{c^2}{\omega^2 z^2}\right]. \label{eq:Fofz}
\end{equation}
The $j=0$ term has been removed from the first sum; this exclusion
of the electron self-field is the explicit renormalization required
to avoid an infinite result.

$\Sigma_P$ depends on an axial offset but not a radial one because
of the transverse symmetry of two parallel plates. Its imaginary
part---cyclotron damping---has a sawtooth form with sharp teeth
where the frequency corresponds to an integral number of
half-wavelengths between the two endcaps (only the odd integers for
$z=0$ since the even integers have a node there). The real
part---cyclotron frequency shifts---shows peaks at similar
intervals. The nearest such frequency corresponds to eight
half-wavelengths at 154.5~GHz, far enough away that the
parallel-plate contribution is smooth in our experimental region of
interest.

The contribution from the cylindrical wall is
% Form that displays the equation in four rows of one column
%\begin{gather}
%   \Sigma_S(\omega,z,\rho) = -\frac{r_0}{z_0}\sum_{p=1}^\infty \sin^2(\tfrac{p\pi}{2}(\tfrac{z}{z_0}+1)) \label{eq:SigmaS} \\
%       \times\sum_{m=0}^\infty (1+\textrm{sgn}(m)) \left[
%                       \frac{K_m^\prime(\mu_p\rho_0)}{I_m^\prime(\mu_p\rho_0)} R_I(m;\mu_p\rho) \right.\notag \\
%                       + \left(\frac{p\pi c}{2\omega z_0}\right)^2
%                       \left(\frac{K_m(\mu_p \rho_0)}{I_m(\mu_p \rho_0)} R_I(m;\mu_p\rho) \right. \notag \\
%                       -\left.\left.\frac{K_m(\tfrac{p\pi\rho_0}{2 z_0})}{I_m(\tfrac{p\pi\rho_0}{2 z_0})} R_I(m;\tfrac{p\pi\rho}{2 z_0})\right)\right] \notag
%\end{gather}
%
% Form that displays the equation across both columns
\begin{widetext}
\begin{align}
    \Sigma_S(\omega,z,\rho) = -\frac{r_0}{z_0}&\sum_{p=1}^\infty \sin^2(\tfrac{p\pi}{2}(\tfrac{z}{z_0}+1)) \sum_{m=0}^\infty (1+\textrm{sgn}(m))  \label{eq:SigmaS} \\
    &\times \left[
                        \frac{K_m^\prime(\mu_p\rho_0)}{I_m^\prime(\mu_p\rho_0)} R_I(m;\mu_p\rho)                        + \left(\frac{p\pi c}{2\omega z_0}\right)^2
                        \left(\frac{K_m(\mu_p \rho_0)}{I_m(\mu_p \rho_0)} R_I(m;\mu_p\rho)
                        -\frac{K_m(\tfrac{p\pi\rho_0}{2 z_0})}{I_m(\tfrac{p\pi\rho_0}{2 z_0})} R_I(m;\tfrac{p\pi\rho}{2 z_0})\right)\right] \notag
\end{align}
\end{widetext}
with
\begin{gather}
    \mu_p = \sqrt{\left(\frac{p\pi}{2 z_0}\right)^2 - \left(\frac{\omega}{c}\right)^2}, \\
    R_I(m;x) = \frac{m^2}{x^2}I_m(x)^2+I_m^\prime(x)^2 ,    \label{eq:RIdefinition} \\
    \textrm{sgn}(m)=\begin{cases}-1 & \textrm{for}~m<0 \\ 0 & \textrm{for}~m=0 \\ 1 & \textrm{for}~m>0\end{cases}.
                        \label{eq:sgndefinition}
\end{gather}
The sums include modified Bessel functions of the first and second kinds,
\begin{eqnarray}
I_\nu(x)&=&i^{-\nu}J_\nu(ix)\\
K_\nu(x)&=&(\pi/2)[(I_{-\nu}(x)-I_\nu(x))/\sin(\nu\pi)]
\end{eqnarray}
as
well as their derivatives. The $K_m^\prime(\mu_p
\rho_0)/I_m^\prime(\mu_p \rho_0)$ term comes from the boundary
conditions of the TE modes, while the $K_m(\mu_p \rho_0)/I_m(\mu_p
\rho_0)$ term comes from the TM modes. For $\rho\rightarrow 0$, the
$R_I$ functions all go to zero except when $m=1$, when it goes to
1/2. For $z\rightarrow0$, only the odd-$p$ terms survive. Combined,
these limits reproduce Eq.~4.28 of
Ref.~\onlinecite{RenormalizedModesPRA}.

The frequency shifts and damping from individual modes can be seen
by looking at the Bessel functions in the denominators. For a given
$p$, an increasing $\omega$ will eventually cross a threshold at
which $\mu_p$ becomes zero and then imaginary. At that point, we may
use the definition of $I_m(x)$ to substitute
\begin{equation}
    I_m(\mu_p\rho_0) = i^{-m} J_m(\tilde{\mu}_p \rho_0) ,
\end{equation}
where $\tilde{\mu}_p$ is the now-real quantity $i \mu_p$. Since
$J_m(x)$ and $J_m^\prime(x)$ have a number of zeros, after $\omega$
exceeds the $p^\textrm{th}$ threshold the sum has poles that may be
approximated as
\begin{equation}
    \Sigma_S(\omega,z,\rho) \approx \frac{\lmnp^2}{\omega^2-\wmnp^2} \label{eq:singlemodeapprox}.
\end{equation}
For TE modes, the poles occur when $J_m^\prime(\tilde{\mu}_p
\rho_0)$ has a zero; for TM modes, when $J_m^\prime(\tilde{\mu}_p
\rho_0)$ has a zero. That is, there are poles when $\omega =\,\wmnp$
of Eq.~\ref{eq:modefreq}. Expanding the Bessel functions about their
zeros yields mode coupling strengths
\begin{subequations} \label{eq:modecouplingparameters}
\begin{align}
    \textrm{TE:}\quad \lmnp^2 =& \frac{2 r_0 c^2}{z_0 \rho_0^2}\frac{-(1+\textrm{sgn}(m))}{J_m^{\prime\prime}(x_{mn}^\prime)J_m(x_{mn}^\prime)} \\
    &\times \sin^2(\tfrac{p\pi}{2}(\tfrac{z}{z_0}+1)) R_J(m;x_{mn}^\prime\tfrac{\rho}{\rho_0}) \notag \\
    \textrm{TM:}\quad \lmnp^2 =& \frac{2 r_0 c^2}{z_0 \rho_0^2}\frac{1+\textrm{sgn}(m)}{J_m^\prime(x_{mn})^2} \left(\frac{p\pi}{2 z_0}\frac{c}{\omega_{mnp}}\right)^2 \\
    &\times \sin^2(\tfrac{p\pi}{2}(\tfrac{z}{z_0}+1)) R_J(m;x_{mn}\tfrac{\rho}{\rho_0}) \notag ,
\end{align}
\end{subequations}
where $x_{mn}^{(\prime)}$ are the previously mentioned zeros of
the Bessel functions and their derivatives. The entire
radial dependence of the coupling is contained in the $R_J$
function, defined by
\begin{equation}
    R_J(m;x) = \frac{m^2}{x^2}J_m(x)^2-J_m^\prime(x)^2 .
    \label{eq:RJ}
\end{equation}
For zero radius, $R_J$ equals 1/2 if $m=1$ and zero otherwise.

Given the above, we can see that the summation indices $p$ and $m$
in Eq.~\ref{eq:SigmaS} correspond directly to those in the mode
indices $mnp$, and the addition of the $m,p^\textrm{th}$ term of the
sums adds the contributions from all modes of that $m$ and $p$. The
threshold above which $\mu_p$ is imaginary corresponds to the
frequency whose half-wavelength fits between the endcaps $p$ times.
The single-mode approximation of Eq.~\ref{eq:singlemodeapprox} shows
the interaction between the electron and a single mode is that of
two weakly coupled oscillators.

\subsubsection{Using the renormalized calculation}\label{sec:UsingTheRenormalizedCalculation}

The combination of $\Sigma_P$ and $\Sigma_S$ in
Eq.~\ref{eq:renormresult} is the result of the renormalized
calculation. There, we have included cavity dissipation in the from
of a mode $Q$ with the replacement $\omega \rightarrow
(1+i/(2Q))\omega$. It is possible to include different quality
factors for the TE and TM mode classes by using $\QTE$ in the
denominator functions $I_m^\prime(\mu_p \rho_0)$ and $\QTM$
everywhere else~\cite{RenormalizedModesPRA}. It is not possible to
include a $Q$ for each mode separately.

The strength of the renormalized calculation is its removal of the
electron self-energy. It has an important drawback in that the
entire calculation has only four input parameters: $\rho_0$, $z_0$,
$\QTM$, and $\QTE$. It does not allow the input of arbitrary mode
frequencies and $Q$s. If the dominant mode-couplings are to one TE
and one TM mode, then the two mode frequencies and $Q$s can
determine the four input parameters. The addition of a third mode,
however, over-constrains the problem; unless the three modes happen
to have frequencies that correspond to those of an ideal cavity and
two happen to share $Q$s, the renormalized calculation will give an
incorrect result.

In this measurement, three modes have the largest influence on the
electron--cavity coupling: TE$_{127}$, TM$_{143}$, and TE$_{136}$.
Because we can identify the terms in Eq.~\ref{eq:SigmaS} that
correspond to a given mode, we modify the renormalized calculation
to better approximate our observed mode structure. Since the
electron is close to (but not precisely in) the mode center, the two
modes with antinodes at the center (odd $p$) dominate the coupling,
and we set the four input parameters of the renormalized calculation
with the frequencies and $Q$s of TE$_{127}$ and TM$_{143}$. These
input parameters give an ideal frequency and $Q$ for TE$_{136}$,
which we correct to the observed values by subtracting the term in
Eq.~\ref{eq:SigmaS} that includes the contribution from TE$_{136}$,
$-(2r_0/z_0)\sin^2(3\pi(z/z_0+1))[K_1^\prime(\mu_6\rho_0)/I_1^\prime(\mu_6\rho_0)]
R_I(1;\mu_6\rho)$, and adding it back with the observed values.
Although this moves all modes with TE$_{1n6}$ the next-nearest ones
are far away: $\omega_{\textrm{TE}126}/(2\pi)\approx129$~GHz and
$\omega_{\textrm{TE}146}/(2\pi)\approx169$~GHz, while
$\omega_{\textrm{TE}136}/(2\pi)\approx146$~GHz. If this were a
concern, we could do a similar subtraction with the single-mode
approximation of Eq.~\ref{eq:singlemodeapprox}, but at the cost of
artifacts of uncanceled higher-order terms.

\subsubsection{Coupling modulated by axial motion}\label{sec:axialamplmodecoupling}

When the SEO is running, the axial motion through the cavity mode
field will modulate the coupling at $\omega_z$. This motion is
present, for example, during cyclotron or spin state-detection; it
is not present during cyclotron or anomaly excitation, the relevant
period for any systematic shift in $g/2$. Since measurements of the
cyclotron damping rate involve continuous detection of the cyclotron
state waiting for a decay, we must account for this modulation and
any amplitude-dependence of the damping rate when using measurements
of $\gc$ to determine the input parameters to the $g/2$ cavity
corrections. This accounting is a nontrivial task, which is
intractable for the full renormalized calculation but doable for the
single-mode coupling of Eq.~\ref{eq:singlemodeapprox} provided that
the axial amplitude $A$ is much lower than a quarter-wavelength of
the mode's axial standing wave ($A\ll z_0/p$), a condition met for
typical $A$.

All of the $z$-dependence in the mode-coupling parameters $\lmnp$ of
Eq.~\ref{eq:modecouplingparameters} comes in a single sine function:
\begin{equation}
    \lmnp = \sin(\tfrac{p\pi}{2}(\tfrac{z}{z_0}+1)) \tilde{\lambda}_\textrm{M} ,
    \label{eq:lambdatildedef}
\end{equation}
where we define $\tilde{\lambda}_\textrm{M}$ to be the
non-$z$-dependent part of the coupling. For an electron offset $z$
from the center of the modes and oscillating at frequency $\omega_z$
with amplitude $A \ll z_0/p$, we may expand the mode
axial-dependence in terms of axial harmonics,
\begin{equation}
    \sin(\tfrac{p\pi}{2}(\tfrac{z+A\cos(\omega_z t)}{z_0}+1)) = \sum_{j=0}^\infty f_j(z,A) \cos(j\omega_z t) ,
    \label{eq:axialharmonicexpansion}
\end{equation}
where the $f_j(z,A)$ are functions of the axial offset and
amplitude. The first three are
\begin{subequations}\label{eq:axialharmonicterms}\begin{align}
    f_0(z,A)&=\sin(\tfrac{p\pi}{2}(\tfrac{z}{z_0}+1))\left[1-\left(\frac{p\pi A}{4z_0}\right)^2+O(A^4)\right] \\
    f_1(z,A)&=\cos(\tfrac{p\pi}{2}(\tfrac{z}{z_0}+1))\left[\frac{p\pi A}{2z_0}- O(A^3)\right] \\
    f_2(z,A)&=\sin(\tfrac{p\pi}{2}(\tfrac{z}{z_0}+1))\left[-\left(\frac{p\pi A}{4z_0}\right)^2+ O(A^4)\right] .
\end{align}
\end{subequations}

Including this expansion of the axial oscillation in the transverse
equation of motion~\cite[App.~A]{ThesisHanneke} yields an amplitude-dependence to the single-mode
coupling strength as well as a series of axial harmonics to the mode
frequency, $\omega_\textrm{M}$:
\begin{align}
    \Dwc-i\frac{\gc}{2}&=\frac{\tilde{\lambda}_\textrm{M}^2\omega}{2}\sum_{j=0}^\infty f_j(z,A)^2 \label{eq:modeaxialamplitude} \\
        &\times \left[  \frac{1}{\omega^2-(\omega_\textrm{M}-j\omega_z)^2}
        + \frac{1}{\omega^2-(\omega_\textrm{M}+j\omega_z)^2}\right] . \notag
\end{align}
As before, we may include a damping width by substituting $\omega
\rightarrow (1+i/(2Q))\omega$ in the two fractions within the
brackets. Note that taking the $A\rightarrow 0$ limit recovers the
usual single-mode coupling of Eq.~\ref{eq:singlemodeapprox}.

\begin{figure}
    \includegraphics[width=\columnwidth]{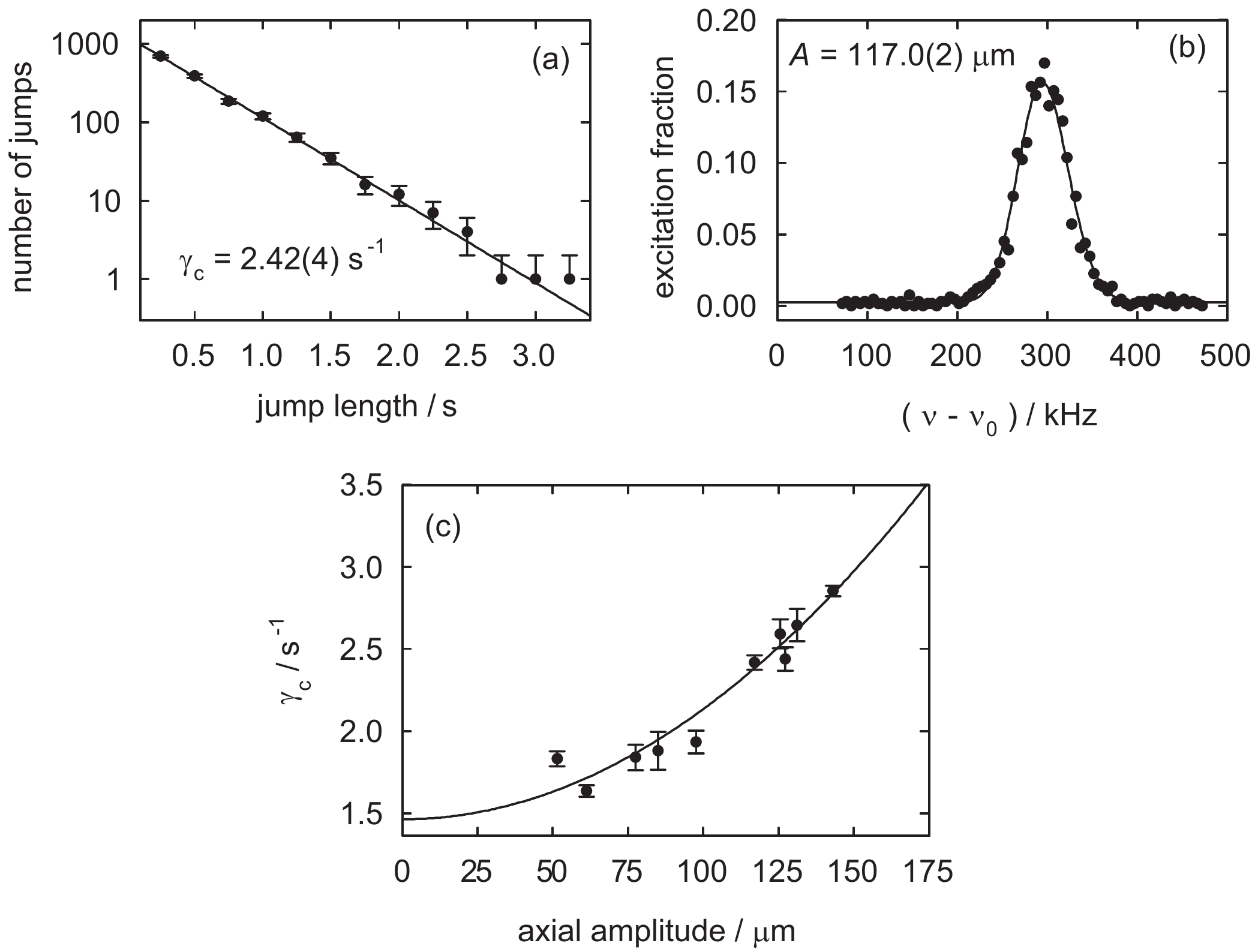}
    \caption[Typical cyclotron damping rate measurement]{Measurement of the cyclotron damping rate at $146.70$~GHz, near the upper sideband of TE$_{136}$. The cyclotron damping rate as a function of axial amplitude (c) extrapolates to the desired lifetime. Each point in (c) consists of a damping rate measured from a fit to a histogram of cyclotron jump lengths (a) as well as an axial amplitude measured from a driven cyclotron line (b).} \label{fig:dampingmeasurement}
\end{figure}

\subsection{Single-electron mode detection} \label{sec:SingleElectronModeDetection}

Using the above descriptions of electron--cavity coupling, we measure the cyclotron damping rate as a function of
cyclotron frequency and of position in the trap to determine the location and $Q$ of the three closest coupled modes
and to characterize the alignment of the electrostatic and mode centers. Since we are able to perform a QND measurement
on the cyclotron state, measuring the cyclotron damping rate simply consists of making many (typically hundreds of)
jumps and fitting the distribution of jump lengths to a decaying exponential with time-constant $\gc^{-1}$, as in
Fig.~\ref{fig:dampingmeasurement}a. Because the axial motion is self-excited during detection, we must measure the
damping rate as a function of amplitude. The amplitude-dependence goes as even powers of $A$
(Eqs.~\ref{eq:axialharmonicterms} \& \ref{eq:modeaxialamplitude}), and since the amplitude is much less than a
quarter-wavelength for the relevant modes, terms of higher-order in $A$ get progressively smaller, allowing
approximation as a quadratic function,
\begin{equation}
    \gamma(A) = \gamma_0 + \gamma_2 A^2 .
\end{equation}

We measure the axial amplitude with the driven cyclotron lineshape of Eq.~\ref{eq:drivenlineshape}, where the driven
axial motion in the magnetic bottle causes the electron to see a higher average magnetic field, resulting in a
cyclotron frequency shift as in Fig.~\ref{fig:dampingmeasurement}b. Figure~\ref{fig:dampingmeasurement}c shows an
example of the measured damping rate as a function of amplitude close to the upper axial sideband of TE$_{136}$
($\nu_{\textrm{TE}136}+\nu_z$). It displays a large amplitude-dependence in $\gc$ because of the proximity of $\nucb$ and
the sideband, which becomes more prominent as larger axial oscillations increase the modulation of the mode coupling.

After repeating such measurements at many cyclotron frequencies, we amass two sets of data, the zero-amplitude
cyclotron decay rates $\gamma_0$ and the quadratic amplitude-dependence coefficients $\gamma_2$, as functions of
$\nucb$. From these data, we extract the frequencies of the three nearest coupled modes: TE$_{127}$, TE$_{136}$, and
TM$_{143}$. Since the lifetimes are heavily $Q$-dependent, we must include three $Q$s, bringing the total number of fit
parameters to six.

The renormalized calculation provides the model for fitting the zero-amplitude cyclotron decay rates. As described
above, we use its four free parameters to determine the frequencies and $Q$s of TE$_{127}$ and TM$_{143}$ and add the
TE$_{136}$ parameters by subtracting the TE$_{1n6}$ term in the renormalized calculation for the calculated frequency
and $Q$ and adding it back in with the frequency and $Q$ as fit parameters.

For the quadratic-amplitude-dependence data, we use the $A^2$ term in the single-mode coupling expansion of
Eq.~\ref{eq:modeaxialamplitude} to write
% Form that displays the equation as five lines in one column
%\begin{align}
%   \gamma_2 =& -2\sum_\textrm{M}~\textrm{Im}\left\{\frac{\tilde{\lambda}_\textrm{M}^2\omega}{2} \right. \label{eq:gamma2} \\
%   &\times \left[
%           \sin^2(\tfrac{p\pi}{2}(\tfrac{z}{z_0}+1)) \left(\frac{p\pi}{4 z_0}\right)^2 \frac{-4}
%           {\omega^2+i\frac{\omega\omega_\textrm{M}}{Q_\textrm{M}}-\omega_\textrm{M}^2} \right. \notag \\
%       &~+ \cos^2(\tfrac{p\pi}{2}(\tfrac{z}{z_0}+1)) \left(\frac{p\pi}{2z_0}\right)^2 \notag \\
%       &~~\times\left( \frac{1}{\omega^2+i \frac{\omega(\omega_\textrm{M}-\omega_z)}{Q_\textrm{M}} -(\omega_\textrm{M}-\omega_z)^2}
%           \right. \notag \\
%       &~~~+\left.\left.\left. \frac{1}{\omega^2+ i\frac{\omega(\omega_\textrm{M}+\omega_z)}{Q_\textrm{M}} -(\omega_\textrm{M}+\omega_z)^2}\right)
%           \right]\right\} , \notag
%\end{align}
%
% Form that displays the equation across both columns
\begin{widetext}
\begin{align}
    \gamma_2 = -2\sum_\textrm{M}~\textrm{Im}\left\{\frac{\tilde{\lambda}_\textrm{M}^2\omega}{2}\right.
     &\left[
            \sin^2(\tfrac{p\pi}{2}(\tfrac{z}{z_0}+1)) \left(\frac{p\pi}{4 z_0}\right)^2 \frac{-4}
            {\omega^2+i\frac{\omega\omega_\textrm{M}}{Q_\textrm{M}}-\omega_\textrm{M}^2}+ \cos^2(\tfrac{p\pi}{2}(\tfrac{z}{z_0}+1)) \left(\frac{p\pi}{2z_0}\right)^2 \right.\label{eq:gamma2} \\
    &\left.\left. \qquad\times
        \left( \frac{1}{\omega^2+i \frac{\omega(\omega_\textrm{M}-\omega_z)}{Q_\textrm{M}} -(\omega_\textrm{M}-\omega_z)^2}
        +\frac{1}{\omega^2+ i\frac{\omega(\omega_\textrm{M}+\omega_z)}{Q_\textrm{M}} -(\omega_\textrm{M}+\omega_z)^2}\right)
            \right]\right\} , \notag
\end{align}
\end{widetext}
where the sum is over the three modes of interest.

We form a $\chi^2_{\gamma_0}$ using the zero-amplitude data and the renormalized calculation and a $\chi^2_{\gamma_2}$
using the quadratic-amplitude-dependence data and Eq.~\ref{eq:gamma2}. Since the electron is close to centered axially
(we discuss the offset below and include the measured offset in the fits), the zero-amplitude data is more sensitive to
the two modes with central antinodes (TE$_{127}$ and TM$_{143}$) and the amplitude-dependence data to the sidebands of
TE$_{136}$, which has a central node. The fit consists of minimizing the two $\chi^2$s; though the proper weighting of
the two is not clear \textit{a priori}, it makes little difference to the results.

%\begin{figure}
%   \includegraphics[width=\columnwidth]{lifetimefit}
%   \caption{Lifetime data with fit. Each cyclotron frequency has a pair of points, one for the zero-axial-amplitude damping rate and one for the quadratic amplitude-dependence, that come from a lifetime-versus-amplitude fit as in Figure~\ref{fig:dampingmeasurement}c.} \label{fig:lifetimefit}
%\end{figure}

Figure~\ref{fig:cavityresults}b\&c display the lifetime data and fits, and Table~\ref{tbl:lifetimeparameters} lists the
results and compares them to those from the parametric mode maps. The two independent methods should agree but do not.
When calculating the cavity shifts, we assign uncertainties large enough to include both results for the mode
frequencies (see Table~\ref{tbl:cavityparameters}). For the mode $Q$s, to which the cavity shifts are largely insensitive, we use the results from the lifetime fits because of the strong-coupling ambiguity in the parametric mode
map values.

\begin{figure}
    \includegraphics[width=\columnwidth]{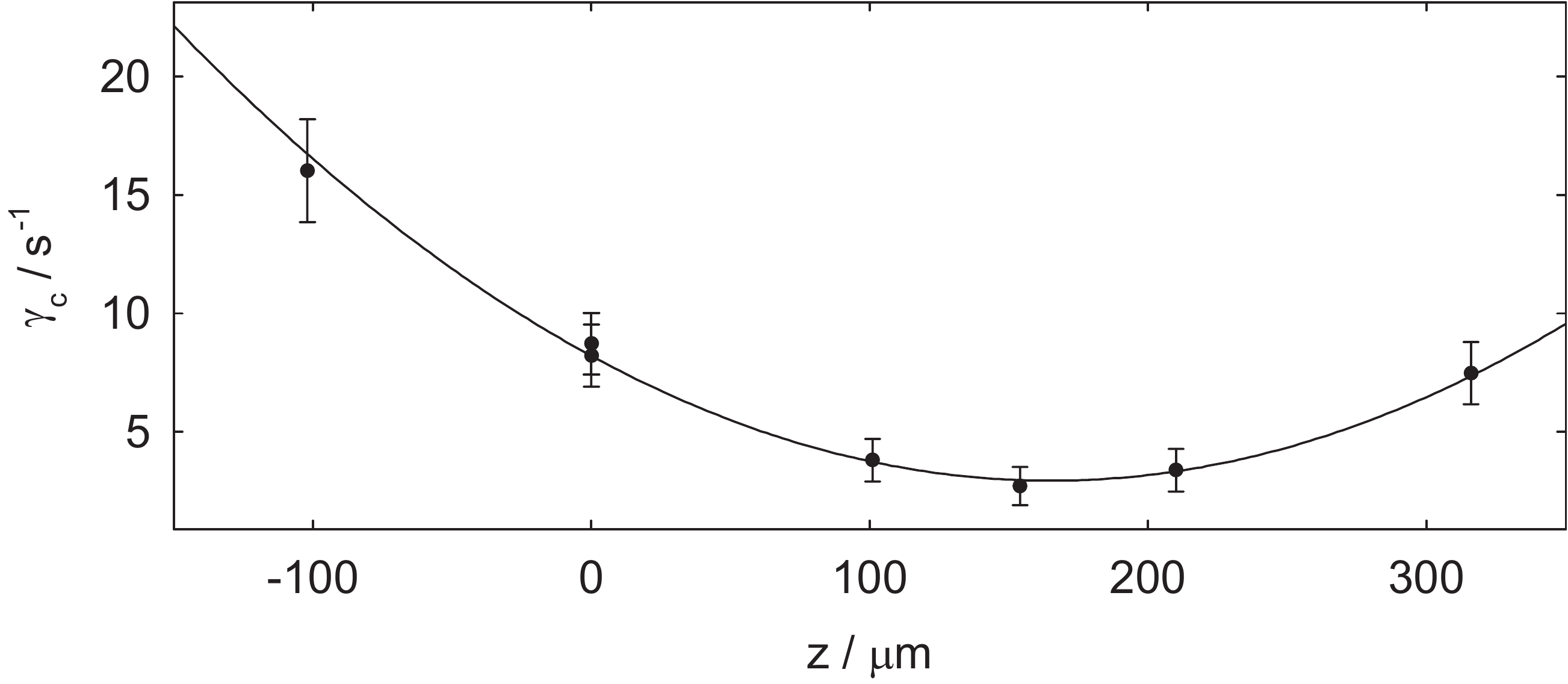}
    \caption{Measurement of the offset between the electrostatic and mode centers. Here, $z$ refers to the position relative to the electrostatic center, and the minimum in the cyclotron damping rate corresponds to the mode center.} \label{fig:axialmisalignment}
\end{figure}

\subsubsection*{Electron position in the cavity}

Knowledge of the electron position relative to the cavity modes is important for calculating the electron--mode
coupling and thus the cyclotron frequency shift. We determine the axial misalignment by measuring the cyclotron damping
rate as a function of $z$, using antisymmetric endcap potentials to move the electron along the trap
axis~\cite{CylindricalPenningTrap}. Figure~\ref{fig:axialmisalignment} plots such a measurement with the cyclotron
frequency tuned midway between TE$_{136}$ and its upper sideband---close enough to TE$_{136}$ to give a large
$z$-dependence but far enough detuned that the cyclotron lifetime is long enough to see single excitations. Since the
mode-couplings are even functions of $z$, the damping rate should go as $z^2$ with an extremum at $z=0$ for the modes.
Figure~\ref{fig:axialmisalignment} shows this extremum, which fits to a misalignment of $165(4)~\mu$m. In addition,
since the extremum is a minimum, the nearest coupled mode must have a node at the mode center (even $p$), demonstrating
that the TE$_{127}$/TE$_{136}$ identification in Fig.~\ref{fig:modemap} is correct.

The cause of this offset is not known, though the consistent presence of TE$_{136}$ in the parametric mode maps
suggests that it is stable. Measurement of a third trap ``center,'' the minimum of the magnetic bottle, agrees with the
electrostatic center. Attempts to model the observed offset as asymmetric spacing between electrodes does not yield a
convincing explanation and more exotic trap deformations such as a tilted endcap or a compensation electrode that
protrudes slightly into the cavity are difficult to model. Because both the parametric mode maps and
Fig.~\ref{fig:axialmisalignment} indicate an offset, we include it when calculating the cavity shifts. We build our
confidence in this procedure by measuring $g/2$ at four cyclotron frequencies with different cavity shifts and showing
the agreement between the predicted and measured shifts.

We estimate the radial alignment of a single electron by tuning its cyclotron frequency into resonance with three modes
that have nodes at the radial center, i.e., have $m\neq1$, and comparing the measured cyclotron damping rate to that
predicted by the renormalized model with $\rho=0$. Since the $m\neq1$ modes do not couple to a radially-centered
electron, a measured cyclotron damping rate that is faster than the calculated damping rate could indicate a radial
misalignment. For the cases where we observe such a discrepancy, we use the full, $\rho$-dependent renormalized
calculation to estimate the range of radial offsets that could explain the observed damping rates (all calculations
include the axial offset of the previous paragraphs). In each of the three cases (TE$_{035}$, TM$_{027}$, TE$_{043}$),
we measure damping rates close to that predicted for $\rho=0$, and we set the limit $\rho < 10~\mu$m.

\subsection{Cavity-shift results}

\begin{table}[b]
        \caption{Parameters used in calculating the cavity shifts. For comparison, we include earlier estimates from the 2006 measurement, which used the same trap cavity.} \label{tbl:cavityparameters}
        \begin{ruledtabular}
      \begin{tabular}{rc|cc}
                & & 2006                                                                                                                                                                 & this \\
                & & measurement~\cite{HarvardMagneticMoment2006} & measurement \\
                \hline
                \multirow{2}{*}{TE$_{127}$} & $\nu_c$\,/\,GHz & 146.350(200) & 146.309(27) \\
                & $Q$ & $> 500$ & 4900(300) \\ \cline{2-2}
                \multirow{2}{*}{TE$_{136}$} & $\nu_c$\,/\,GHz & --- & 146.428(15) \\
                & $Q$ & --- & 4800(200) \\ \cline{2-2}
                \multirow{2}{*}{TM$_{143}$} & $\nu_c$\,/\,GHz & 151.900(200) & 151.832(37) \\
                & $Q$ & $> 500$ & 1270(70) \\ \cline{2-2}
                electrostatic & z\,/\,$\mu$m & 0 & 165(4) \\
                offset & $\rho\,/\,\mu$m & 0 & $< 10$ \\
            \end{tabular}
        \end{ruledtabular}
\end{table}

\begin{figure}
    \centering
    \includegraphics[width=\columnwidth]{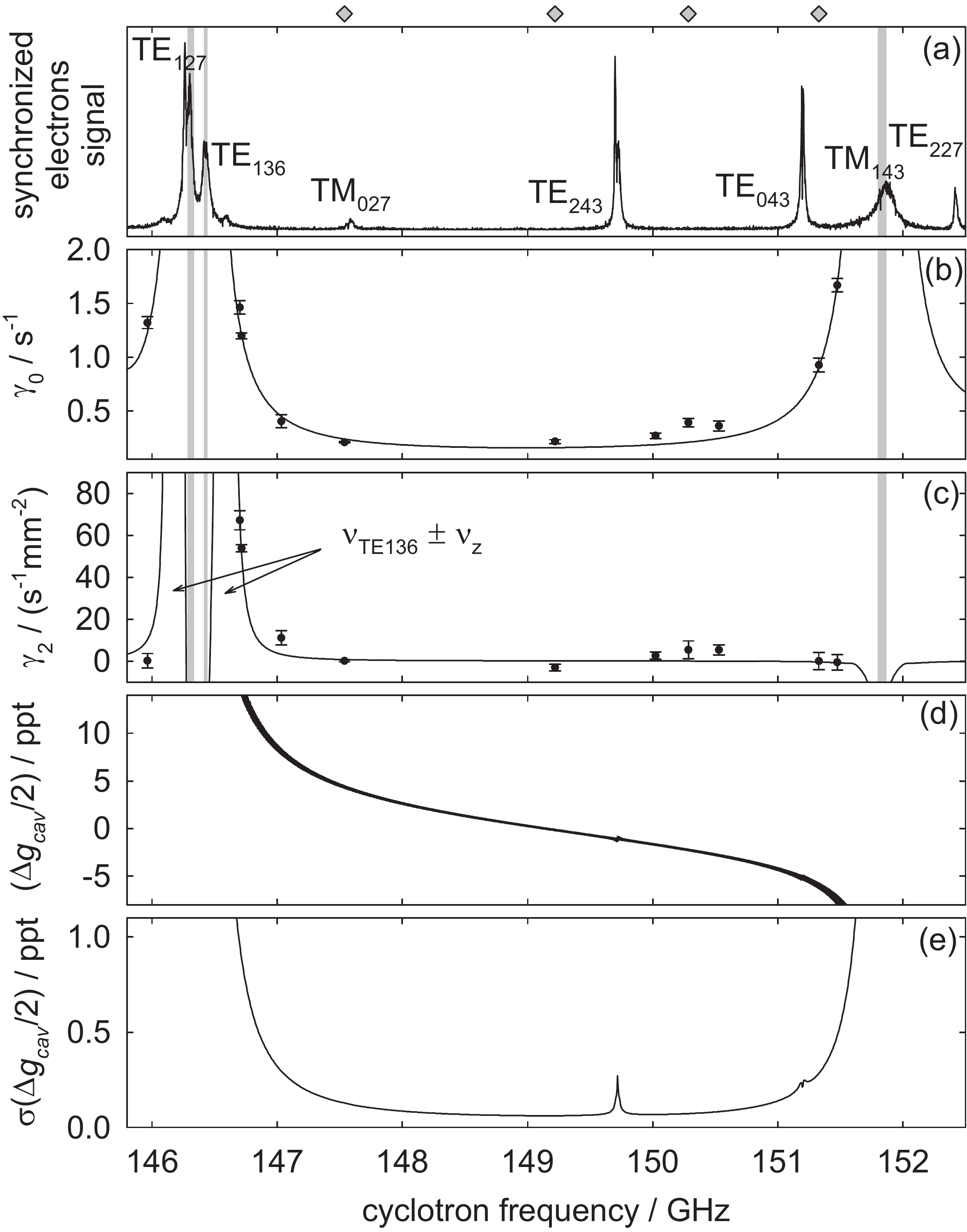}
    \caption{Cavity shift results come from synchronized electrons (a) and from direct measurements with one electron of $\gc$ (b) and its dependence on axial amplitude (c). Together, they provide uncertainties in the frequencies of coupled cavity modes (gray) that translate into an uncertainty band of cavity shifts $\Delta g_{cav}/2$ (d) whose half-width, i.e., the cavity shift uncertainty, is plotted in (e). The diamonds at the top indicate the cyclotron frequencies of the four $g/2$ measurements.} \label{fig:cavityresults}
\end{figure}

We calculate the cavity shifts and uncertainties from the mode parameters and their uncertainties
(Table~\ref{tbl:cavityparameters}) via the renormalized calculation. The cavity shifts are independent of mode $Q$ for
cyclotron frequencies with relative detunings $(\wcb-\omega_\textrm{M})/\omega_\textrm{M}\gg1/(2Q_\textrm{M})$.
Because all our $g/2$ measurements meet this requirement easily, the cavity shift uncertainty comes only from the
errors in mode frequencies. Because of the axial offset, mode TE$_{136}$ affects the shifts, and we calculate with this
mode at its observed frequency using the mode-moving technique described in Sec.~\ref{sec:UsingTheRenormalizedCalculation}. Because fits to the parametric mode
maps and to the single-electron lifetime data yield slightly different frequencies for the three nearest coupled modes
(see Table~\ref{tbl:lifetimeparameters}), we assign uncertainties large enough to include both. The trap-radius limit
only has a significant effect near two modes, TE$_{243}$ and TE$_{043}$, and we again use the mode-moving technique to
place them at their observed frequencies though it makes little difference to our result because none of our four
measurements of $g/2$ were resonant with either mode. Because it appears in the parametric mode map, we include
TM$_{027}$ in the calculation, although it does not change the result noticeably.

%\begin{figure}
%   \includegraphics[width=\columnwidth]{gvalueuncertaintybands}
%   \caption{Cavity shift results. Uncertainties in the frequencies of the coupled cavity modes (a) translate into an uncertainty band of cavity shifts $\Delta g_{cav}/2$ (b) whose half-width, i.e., the cavity-shift uncertainty, is plotted in (c). The diamonds at the top indicate the cyclotron frequencies of the four new $g/2$ measurements.}
%   \label{fig:gvalueuncertaintybands}
%\end{figure}

Figure~\ref{fig:cavityresults}d\&e display the results of this analysis, and Table~\ref{tbl:guncertainties} shows the
calculated cavity shifts for our four measurements of $g/2$. The shifts span over 10~ppt with uncertainties around 100
times smaller than that range. The lowest uncertainties are below a part in $10^{13}$, over six times smaller than in
our 2006 measurement~\cite{HarvardMagneticMoment2006} and low enough that this systematic uncertainty is no longer a
dominant error in the measurement of $g/2$.

\section{Power shifts} \label{sec:PowerShifts}

We expect neither $\nuab$ nor $\fcb$ to shift with cyclotron or anomaly power, but previous measurements of the
electron $g/2$ at the UW showed unexplained systematic shifts of the cyclotron frequency with both drive
powers~\cite{DehmeltMagneticMoment,VanDyckLossy}. The origin of these power shifts in the UW measurements remains
unknown~\cite{VanDyckLossy}, and extrapolation to zero cyclotron power involved correcting shifts of several ppt in
$g/2$~\cite{DehmeltMagneticMoment,VanDyckLossy}. Estimates comparing our drive power to that used in the UW
measurements suggest that our narrower lines and single-quantum cyclotron technique require drive strengths low enough
that the power shifts are negligible. We have yet to see a power shift in our apparatus, though experimental searches
are time-consuming and the statistical uncertainty in the current search is comparable to our final uncertainty in
$g/2$.

\subsection{Anomaly power shift estimate}

The UW experiment showed anomaly power shifts of several ppb in the anomaly frequency~\cite{VanDyckLossy}. An
off-resonant anomaly drive during cyclotron excitation shifted the cyclotron line by a similar amount, and the two
shifts canceled in the frequency-ratio calculation of $g/2$~\cite{DehmeltMagneticMoment}. The origin of these shifts is
unknown, although experiments with a variable-strength magnetic bottle showed that they increase with the magnitude of
the bottle strength, independent of its sign~\cite{VanDyckLossy}.

Direct comparisons between the anomaly power used in the UW experiment and that used here are difficult because the
experiments use different anomaly excitation techniques. The UW excitations were primarily driven with counterflowing
current loops in split compensation electrodes, while we drive the electron axially through the $z\rho\rhohat$ gradient
of the magnetic bottle. Unlike the current-loop excitation technique, our axial-excitation technique provides a clear
mechanism for an anomaly power shift by increasing the average axial amplitude and, therefore, the average magnetic
field seen by the electron (because of the $z^2\zhat$ part of the magnetic bottle). We estimate this shift below and
expect it to be both smaller than our current precision and canceled in the calculation of $g/2$ by a similar cyclotron
shift from a detuned anomaly drive during cyclotron excitation.

For a driven axial amplitude, $z_a$, the frequency shift from the motion through the magnetic bottle is
\begin{equation}
    \frac{\Delta\omega_a}{\omega_a} = \frac{\Delta\omega_c}{\omega_c} = \frac{B_2}{B}\frac{z_a^2}{2}.
\end{equation}
Were this shift to approach the linewidth, we would be forced to use the driven lineshape of
Eq.~\ref{eq:drivenlineshape} when extracting frequencies. In order to calculate the expected size of the power shift,
we must estimate $z_a$; only amplitudes over 800~nm will produce shifts at the 0.1~ppb level in frequency. We estimate
$z_a$ using two methods: the observed anomaly transition rate and a calibration of the drive voltage. The estimates
give similar amplitudes, and neither predicts anomaly power shifts at our precision.

From Eq.~\ref{eq:unsaturatedprob}, we see that the anomaly transition rate goes as the product of the Rabi frequency
squared $\Omega_a^2$ times the lineshape function $\chi(\omega)$. The anomaly Rabi frequency goes as
$z_a$~\cite{Palmer}, and we can estimate $z_a$ from a typical peak excitation fraction, $P_\textrm{pk}$, using
\begin{align}
    P_\textrm{pk}&=\frac{\pi}{2}T\Omega_a^2\chi(\omega_\textrm{pk}) \\
        &=\frac{\pi}{2}T\left(\frac{g}{2}\frac{e\hbar}{2m}
            B_2 z_a
            \sqrt{\frac{2}{m\hbar(\wcb-\bar{\omega}_m)}}\right)^2
            \chi(\omega_\textrm{pk}). \notag
\end{align}
Because the lineshape function is normalized to unity, its value on-peak in the Lorentzian lineshape limit---the limit
corresponding to the anomaly line---is inversely-proportional to the linewidth
\begin{equation}
    \chi(\omega_\textrm{pk})=\frac{2}{\pi}\left(\frac{2\Dw^2}{\gz}+\gc\right)^{-1}.
\end{equation}
For typical experimental parameters, we must drive to $z_a \approx 100$~nm to achieve a 20\% excitation fraction.

Alternately, we can estimate the driven amplitude based on the rf voltage on the bottom endcap. An endcap driven with
amplitude $V_a$ excites the electron to an amplitude given by~\cite{Review}
\begin{equation}
    z_a = \frac{c_1 d^2}{2 z_0} \left[\left(\frac{\omega_a}{\omega_z}\right)^2-1\right]^{-1} \frac{V_a}{V_R}.
\end{equation}
($V_R$ is the ring electrode potential, and $c_1$, $d$, and $z_0$ are geometric factors equal to approximately 0.78,
3.5~mm, and 3.8~mm in our trap.) We calibrate the drive amplitude using the anomaly-power-induced axial frequency
shifts discussed in Sec.~\ref{sec:ExperimentalRealization}. They indicate 30~dB of attenuation between the anomaly frequency
synthesizer and the trap electrode. This agrees with the attenuation we measure in the drive line during
room-temperature calibrations and seems reasonable given the 20~dB cold attenuator installed at the 1K pot and some
additional loss in the stainless steel semi-rigid coaxial cable. The highest anomaly power used for $g/2$ data was
-16~dBV at the synthesizer, which would attenuate to $V_a = 5$~mV at the bottom endcap and drive the electron to $z_a =
250$~nm.

Driven axial amplitudes around 100--250~nm should only shift the anomaly and cyclotron frequencies at the
1--10~ppt-level, which is far too small to affect the lineshapes (the error in $g/2$ would be lower by 1000 if the
cyclotron and anomaly shifts were uncorrelated, but should be even smaller because the shifts cancel in the frequency
ratio).

\subsection{Cyclotron power shift estimate}

Unlike the anomaly power shifts, the cyclotron power shifts seen in the UW experiments did not cancel in $g/2$ and
added a 1.3~ppt uncertainty to their 1987 result~\cite{DehmeltMagneticMoment}. The shifts appeared as a resonant effect
of unknown origin with a resonant drive shifting the cyclotron line several ppb but a detuned cyclotron drive with the
same power not shifting the anomaly line. Investigations in a trap with a variable-strength magnetic bottle showed that
the shift scaled with $B_2$ in magnitude and sign. In~\cite{VanDyckLossy}, the authors hypothesize that the shift could
have originated in an excitation of the magnetron motion because a typical shift could have been explained by a 10\%
increase of the magnetron radius.

Our cyclotron excitation technique, injecting microwaves into the trap cavity, is similar to that used in the UW
measurements, so we can compare our technique to theirs. Our lower temperature narrows the lines by a factor of ten,
requiring less power to drive transitions. The measured bottle-dependence suggests that our ten-times-stronger magnetic
bottle could cancel the advantage of our narrower lines. The overall shifts should still be reduced because our
single-quantum-jump spectroscopy only needs to excite to the $n=1$ state less than 20\% of the time. At the UW, typical
excitations sustained the electron at energies corresponding to $n\gtrsim4$~\cite{Review,VanDyckLossy}. Naively,
exciting to an average energy of $n=4$ requires 20-times more power than an average energy of $n=0.2$, and this power
reduction alone would reduce several-ppt shifts in $g/2$ below our precision. The relativistic shifts between cyclotron
levels suggest additional power in the UW drives because excitations above $n=1$ involve driving in the exponential
tails of the higher states' resonances. In addition, if the power shift is indeed related to driving a
magnetron--cyclotron sideband, our ten-times-higher magnetron frequency and ten-times-narrower cyclotron lines put the
closest magnetron sideband, which was fewer than 10 linewidths away at the UW, 100 times farther from the cyclotron
resonance.

\subsection{Experimental searches for power shifts}

\begin{figure}
    \centering
    \includegraphics[width=\columnwidth]{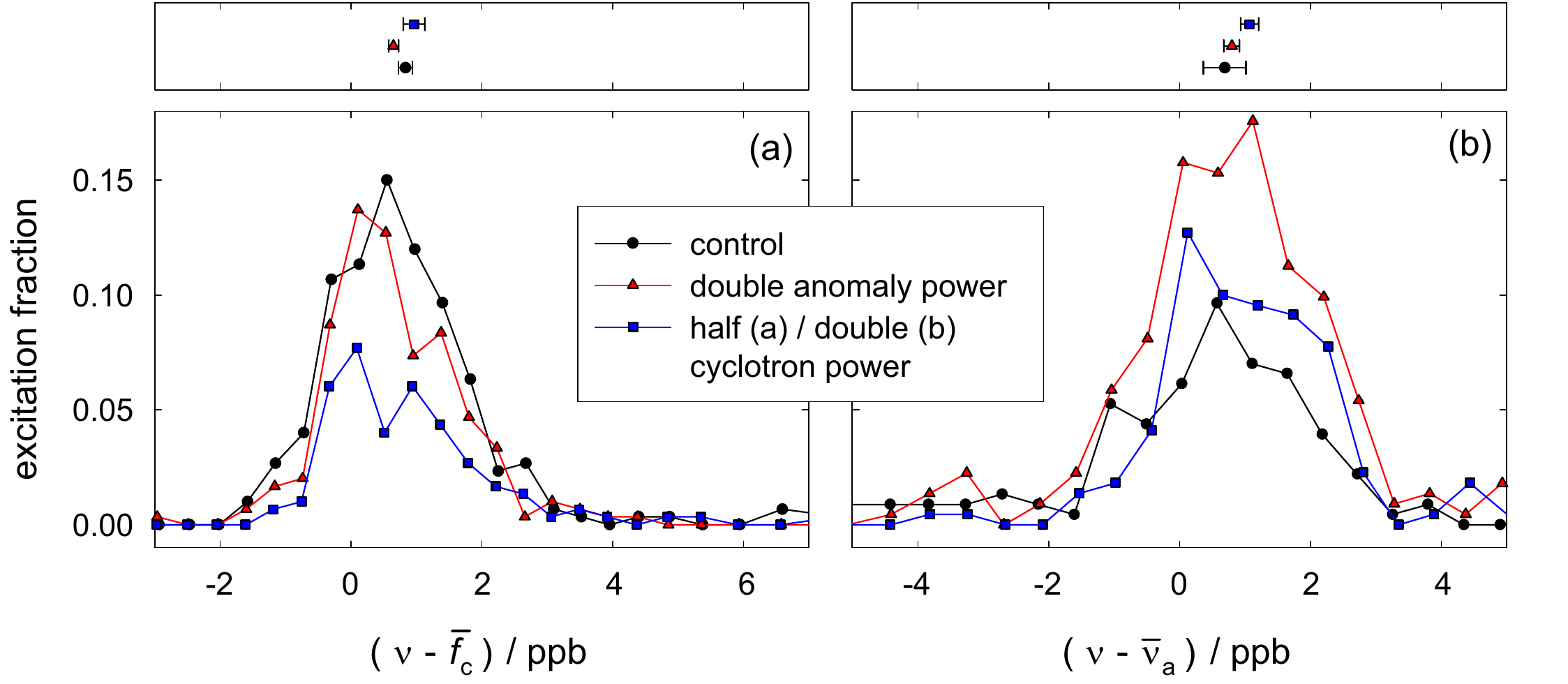}
    \caption{The lower plots show cyclotron (a) and anomaly (b) data taken in search of cyclotron and anomaly power shifts. The upper plots compare the weighted-means of these lines.} \label{fig:powersystematics}
\end{figure}

Although we do not expect any cyclotron or anomaly power shifts of $\nuab$ or $\fcb$, their existence in the UW
measurements makes us proceed with caution and look for them anyway. We examine the shifts of each line individually to
ensure that no systematic effects (even ones that cancel in $g/2$) go unnoticed. We look for a cyclotron frequency
shift by running three cyclotron scans: a control, one with double the detuned anomaly power, and one with half the
cyclotron power (lower to avoid saturation). The scans are interleaved in the same way we interleave cyclotron and
anomaly scans during $g/2$ measurements, alternating single sweeps of each line and including edge-tracking to remove
long-term drifts (see Sec.~\ref{sec:ExperimentalRealization}). The resulting cyclotron lines are shown in
Fig.~\ref{fig:powersystematics}a. We calculate the cyclotron frequency of each line with the weighted-mean method (the
offset from $\fcb$ cancels when subtracting for a frequency shift). Frequency differences between methods are
summarized in Table~\ref{tbl:powersystematics}.

\begin{table}
        \caption{Summary of power-shift searches}\label{tbl:powersystematics}
        \begin{ruledtabular}
      \begin{tabular}{lr}
                \multicolumn{1}{c}{test} & \multicolumn{1}{c}{``shift'' / ppb}\\
                \hline
                    $\fcb$ with double anomaly power            & -0.18\,(13) \\
                    $\fcb$ with half cyclotron power            &  0.13\,(19) \\
                    $\nuab$ with double anomaly power           &  0.11\,(35) \\
                    $\nuab$ with double cyclotron power     &  0.38\,(35) \\
            \end{tabular}
        \end{ruledtabular}
\end{table}

%\begin{table}
%       \caption{Summary of power-shift searches}\label{tbl:powersystematics}
%       \begin{ruledtabular}
%      \begin{tabular}{lr@{.}l@{\,(}l}
%               \multicolumn{1}{c}{test} & \multicolumn{3}{c}{``shift'' / ppb}\\
%               \hline
%                   $\fcb$ with double anomaly power            & -0&18&13) \\
%                   $\fcb$ with half cyclotron power            &  0&13&19) \\
%                   $\nuab$ with double anomaly power           &  0&11&35) \\
%                   $\nuab$ with double cyclotron power     &  0&38&35) \\
%           \end{tabular}
%       \end{ruledtabular}
%\end{table}

To look for anomaly frequency shifts, we run three anomaly scans---a control, one with double the detuned cyclotron
power, and one with double the anomaly power (the control power is low enough that we can double the power without
saturating)---interleaved and normalized via edge-tracking as before. The resulting anomaly lines are shown in
Fig.~\ref{fig:powersystematics}b, which includes the frequencies calculated by the weighted-mean method.
Table~\ref{tbl:powersystematics} summarizes the differences.

The results in Table~\ref{tbl:powersystematics} as consistent with zero. The largest ``shift''
is that of the anomaly frequency with cyclotron power---the only one of the four \textit{not} seen at the UW. The data
of Table~\ref{tbl:powersystematics} suggest that any power shift will be $\lesssim 0.35$~ppb in frequency, which is
consistent with the limits of our prior studies (detailed in Sec.\,6.2 of Ref.~\cite{ThesisOdom}) and with our
expectation of no shift at our current precision. The uncertainties are limited by our ability to resolve the lines in
a timely manner. (The number of nights spent assembling the data in Fig.~\ref{fig:powersystematics} exceeds half the number used
to determine the $g/2$ value.)  The anomaly line in particular requires the time-consuming discrimination between
$\ket{0,\uparrow}$ and $\ket{1,\downarrow}$ after each anomaly pulse, and any search for a systematic shift in the
anomaly frequency multiplies the number of times this must occur. Because we estimate that no power shift should occur
and our experimental searches are limited by our ability to resolve the lines, we apply neither a correction nor any
additional uncertainty from power shifts.

\section{Results and Applications}

\subsection{Most Accurate Determination of the Electron g Value}

\begin{table}[b]
     \caption{Measurements and shifts with uncertainties, multiplied by $10^{12}$.  The cavity-shifted ``$g/2$ raw'' and corrected ``$g/2$'' are offset from our result in Eq.~\ref{eq:g}.}\label{tbl:guncertainties}
    \begin{ruledtabular}
     \begin{tabular}{lr@{\,(}lrrr}
               $\fcb$    & \multicolumn{2}{r}{147.5 GHz} & 149.2 GHz& 150.3 GHz& 151.3 GHz\\
               \hline\hline
               $g/2$ raw &-5.24&0.39)& 0.31\,(0.17) & 2.17\,(0.17)& 5.70\,(0.24) \\
               \hline
                   Cav.\ shift    & 4.36&0.13) & -0.16\,(0.06) & -2.25\,(0.07) & -6.02\,(0.28) \\

                   Lineshape & \multicolumn{5}{r}{}\\
                   \multicolumn{2}{l@{\,(}}{~~correlated}        & 0.24) & (0.24) & (0.24) & (0.24) \\
                   %\multicolumn{2}{l}{~~(uncorrelated)}&&&\\
                   \multicolumn{2}{l@{\,(}}{~~uncorrelated}      & 0.56) & (0.00) & (0.15) & (0.30) \\
                             \hline
                   $g/2$ & -0.88&0.73) & 0.15\,(0.30) & -0.08\,(0.34) & -0.32\,(0.53) \\
    \end{tabular}
    \end{ruledtabular}
\end{table}

\begin{figure}
    \centering
    \includegraphics[width=\columnwidth]{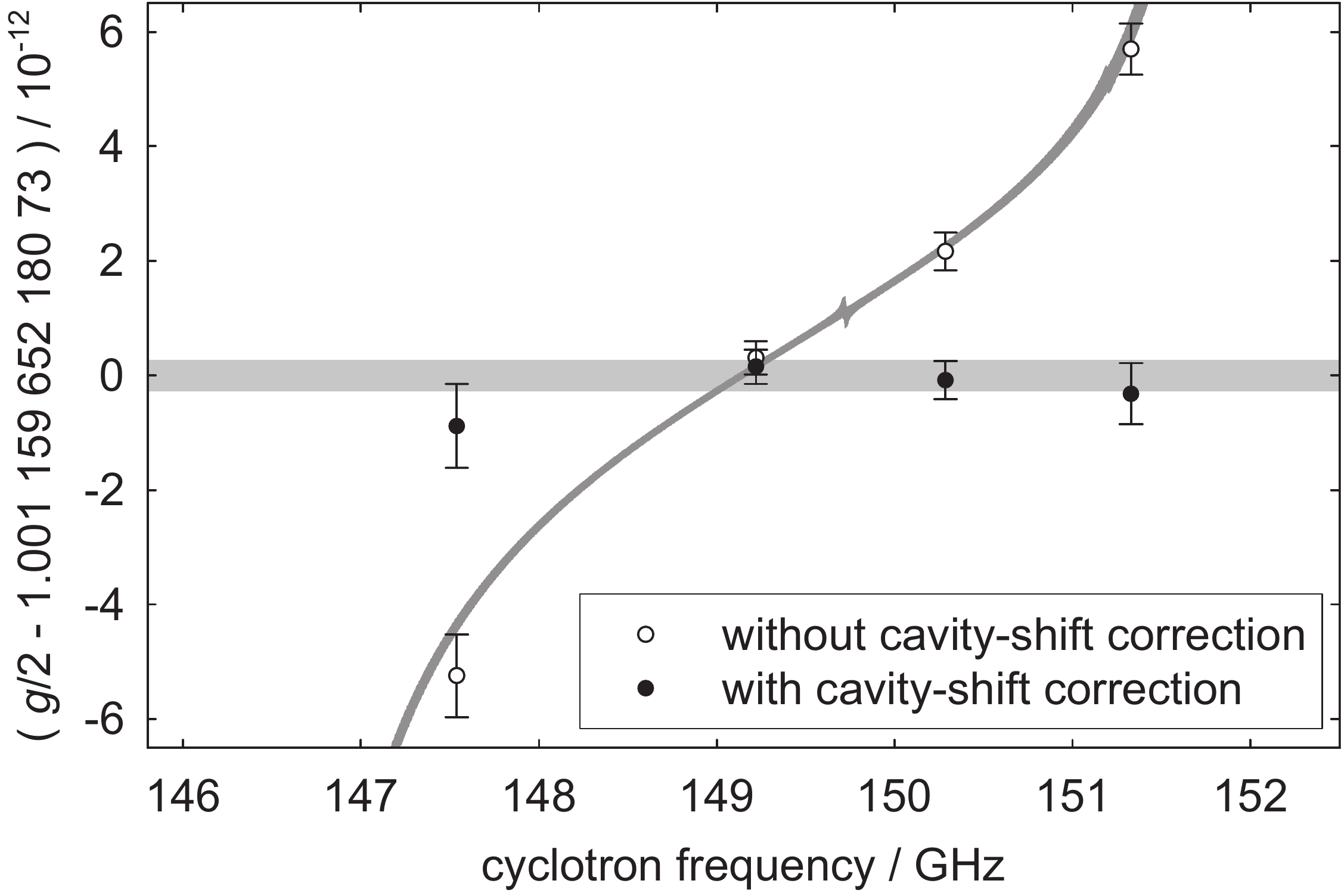}
    \caption{Four measurements of $g/2$ without (open) and with (filled) cavity-shift corrections. The light gray uncertainty band shows the average of the corrected data. The dark gray band indicates the expected location of the uncorrected data given our result in Eq.~\ref{eq:g} and including only the cavity-shift uncertainty.} \label{fig:newgvaluedataandresult}
\end{figure}

The result for the electron magnetic moment in Bohr magnetons,
\begin{equation}
    g/2 = 1.001~159~652~180~73~(28)\quad[0.28~\textrm{ppt}],
    \label{eq:g}
\end{equation}
comes from the weighted average of the four measurements with uncorrelated and correlated uncertainties combined
appropriately. The result has 2.7 and 15 times lower uncertainty than the 2006 and 1987 measurements and 2\,300 times
lower uncertainty than has been achieved for the heavier $\mu$ lepton~\cite{gMuon2006}. Table~\ref{tbl:guncertainties}
summarizes the measured values, shifts, and uncertainties for the four separate measurements of $g/2$. The
uncertainties are lower for measurements with smaller cavity shifts and narrower linewidths, as might be expected. We
no longer quote uncertainties for variations of the power of the $\nuab$ and $\fcb$ drives; although power-shifts
appeared  in the 1987 measurement~\cite{DehmeltMagneticMoment,VanDyckLossy}, our narrower lines and single-quantum
cyclotron technique require much lower drive powers and we estimate---and check experimentally---that they are no
longer important.

\subsubsection*{Relation of 2006 and 2008 measurements}

The 2008 measurement \cite{HarvardMagneticMoment2008} is an independent measurement that is consistent with the 2006 measurement \cite{HarvardMagneticMoment2006}. Essentially the same apparatus is used to make a fresh data set.  However, the apparatus is now better understood and both the measurement and analysis procedures are significantly improved.     For example, the electron is used as a relative magnetometer to allow many data sets, measured on different days, to be combined.  This gives lineshape curves with a signal-to-noise good enough to compare their shape with theoretical expectations.  (Previously, each day's measurements were combined to get a $g$ value, and the values from different days were averaged.)

The biggest reduction in the uncertainty in our second measurement comes from a better understanding of cavity shifts. Two independent probes of the cavity mode structure allow identification of nearly all modes and the quantification of an offset between the effective center of the trap for the radiation modes and for the electrostatic quadrupole potential.    By measuring $g/2$ at four magnetic fields with cavity shifts spanning over 30 times our final uncertainty, we precisely test this once-dominant uncertainty, and demonstrate that we can assign an uncertainty that is much  smaller than that estimated for our first measurement.

Retroactively applying the improved understanding and the modified analysis developed for the second measurement to the first would require starting from the  raw data.
 Improvements in measurement methods cannot be retroactively implemented, of course. We thus believe that the 2008 measurement should be regarded as superseding the 2006 measurement rather than trying to average the two measurements, which would only insignificantly change the value of $g/2$  in the second digit of the 2008 uncertainty.  The correlations between the possible systematic uncertainties that limit the two measurements has neither been studied nor reported carefully enough to allow an appropriate averaging of the  two measurements.

\subsection{Most Accurate Determination of $\alpha$} \label{sec:FineStructureConstant}

The new measurements of the electron $g/2$ determine the fine structure constant about 12 times more accurately than the next-most-precise method
(Fig.~\ref{fig:AlphaComparison}). The relationship between $g/2$ and $\alpha$ has been summarized in detail in \cite{Alpha2006fixed}, with the final value updated in
\cite{HarvardMagneticMoment2008}. Here we give only the results and a brief summary.

Within the standard model of particle physics the electron $g/2$ is related to $\alpha$ by
\begin{align}
\frac{g}{2} =1 &+ C_2\left(\frac{\alpha}{\pi}\right) +C_4\left(\frac{\alpha}{\pi}\right)^2 +C_6\left(\frac{\alpha}{\pi}\right)^3
+C_8\left(\frac{\alpha}{\pi}\right)^4 \notag\\
&+...+a_{\mu,\tau}+a_{\text{hadronic}}+a_{\text{weak}}. \label{eq:gAlpha}
\end{align}
The leading contribution to $g/2$ is the $1$ that is predicted for a Dirac point particle.  Vacuum fluctuations modify the interaction of the
electron with the magnetic field, increasing the effective magnetic moment of the electron by approximately one part per thousand.  This addition is
described by the infinite QED series in powers of $\alpha/\pi$, with coefficients $C_n$ determined by $n$-vertex QED calculations for the interaction
of electrons and photons. The first three coefficients ($C_2$, $C_4$, and $C_6$) have all been calculated exactly. A substantial numerical
calculation has determined $C_8$, and a numerical calculation of $C_{10}$ is underway. A related series involving the $\mu$ and $\tau$ leptons
yields a small contribution, $a_{\mu,\tau}$. Much smaller hadronic and weak contributions, $a_\text{hadronic}$ and $a_\text{weak}$, have been
calculated accurately enough that they do not add uncertainty at the current level of precision. References to the most recently calculated values
are provided in \cite{Alpha2006fixed}.

The fine structure constant is determined from the measured $g/2$ by solving Eq.~\ref{eq:gAlpha} for $\alpha$ to obtain
\begin{eqnarray}
\alpha^{-1} &=& 137.035 \, 999 \, 084 \, (33)\,(39)~~[0.24~\rm{ppb}]\,[0.28~\rm{ppb}],\nonumber \\
&=& 137.035 \, 999 \, 084 \, (51)~~~~~~~~[0.37~\rm{ppb}]. \label{eq:HarvardAlpha2008}
\end{eqnarray}
The first line shows experimental (first) and theoretical (second) uncertainties that are nearly the same. The uncertainty in $\alpha$ is now limited
a bit more by the need for a higher-order QED calculation (underway~\cite{RevisedC8}) than by the measurement uncertainty in $g/2$.

In more detail, the theory uncertainty contribution to $\alpha$ is divided as $(12)$ and $(37)$ for $C_8$ and $C_{10}$.  It should decrease when a
calculation underway \cite{RevisedC8} replaces the crude estimate $C_{10}=0.0 \,(4.6)$ \cite{CODATA2002,Alpha2006fixed}.  The $\alpha^{-1}$ of
Eq.~\ref{eq:HarvardAlpha2008} will then shift by $2\alpha^3 \pi^{-4} C_{10}$, which is $8.0\, C_{10} \times 10^{-9}$.  A change $\Delta_8$ in the
calculated $C_8=-1.9144\,(35)$ would add $2\alpha^2\pi^{-3}\Delta_8$.

The independent methods for determining $\alpha$ that come closet to our accuracy  are the ``atom-recoil'' measurements, so called because their
uncertainty is limited by measurements of recoil velocities in Rb and Cs atoms. They rely on many experiments, including the measured Rydberg
constant~\cite{Hydrogen1s2s2000,FitDeterminesRydberg}, the Rb or Cs mass in amu~\cite{PritchardMassRatios1999}, and the electron mass in
amu~\cite{VanDyckElectronMass,ElectronMassFromBoundg}. The needed $h/M[\textrm{Rb}]$ comes from a measurement of the recoil of a Rb atom in an
optical lattice~\cite{AlphaRbPRL2006,AlphaRbPRA2006,AlphaRb2008}. The needed $h/M[\textrm{Cs}]$ comes from an optical measurement of the Cs D1
line~\cite{Tanner2006} and the ``preliminary'' recoil shift for a Cs atom in an atom interferometer~\cite{ChuAlphaPreliminary2002}.  Although these
determinations of $\alpha$ have an uncertainty that is currently 12--22 times larger than ours, improvements are expected in experiments that are
underway \cite{ImprovedAlphaExpected1,ImprovedAlphaExpected2}.

\subsection{Most Precise Test of QED}

The most stringent test of QED, to the highest order in $\alpha/\pi$, comes from comparing the measured $g/2$ to the value calculated using
Eq.~\ref{eq:gAlpha} using the best available value of $\alpha$ that is not determined from the electron $g$. Our latest $g$, compared to
Eq.~\ref{eq:gAlpha} with $\alpha(\rm{Rb})$, gives a difference \cite{Alpha2006fixed,AlphaRb2008}
\begin{equation}
|\delta g/2| < 8 \times 10^{-12}.
\end{equation}  The good agreement testifies to the
remarkable success of QED. The prototype of modern physics theories is thus tested far more stringently than its inventors ever envisioned
\cite{DysonLetter}.

The latest $g/2$ measurement is now accurate enough to allow a 10--20 times more stringent test of QED, should a comparable-accuracy measurement of $\alpha$ become available.  We thus strongly emphasize the compelling need for greatly improved independent measurements of $\alpha$.

\subsection{Limits on Electron Substructure}

The same comparison of the measured $g/2$ and the value calculated from Eq.~\ref{eq:gAlpha} using the best available independent $\alpha$ probes the
internal structure of the electron \cite{Alpha2006fixed,BrodskyDrellElectronSubstructure}.  A composite electron is constrained to have constituents with
a mass $m^*>m /\sqrt{\delta g/2}=180~\rm{GeV}/c^2$, corresponding to an electron radius $R<1 \times 10^{-18}$ m.

If this test was limited only by the experimental uncertainty in $g/2$ (i.e.\ if a much better independent $\alpha$ becomes available) then we could
set a limit $m^*>1$ TeV. These high energy limits seem somewhat remarkable for an experiment carried out at $100$ mK. However, a search for a
contact interaction in electron-positron collisions at LEP sets a more stringent limit, $m^*>10$ TeV \cite{LepElectronStructureLimit}.

\subsection{Test of CPT Invariance with Leptons}

Already the most precise test of CPT invariance with a lepton system comes from comparing the measured magnetic moment of the positron and the
electron \cite{DehmeltMagneticMoment}.  A new measurement underway at Harvard aims to improve the sensitivity of this test by a factor of 15 or more,
by applying the demonstrated new electron methods to a positron.

\subsection{Application to Dark Matter}

The comparison of the measured $g/2$ and the value calculated from Eq.~\ref{eq:gAlpha} using the best available independent $\alpha$ is also relevant
to one model that attempts to explain dark matter. The measured $g/2$ is accurate enough to allow the discovery of, or to rule out, proposed
dark-matter particles with a mass that is close to the electron mass \cite{LightDarkMatterLimit}, if and when a more accurate independent measurement
of $\alpha$ becomes available.

\section{Outlook}

The new $g/2$ prepares the way for further tests of the standard model, pending the availability of an independent
$\alpha$ at the uncertainty reported here. In addition, the techniques used to measure the electron $g/2$ clear the way
for a series of new measurements, some of which have already begun.

First, measuring the positron $g/2$ to the same precision would improve upon the most stringent lepton CPT
test~\cite{DehmeltMagneticMoment} and constrain possible violations of Lorentz
invariance~\cite{TestingCptWithMagneticMoments}. Except for the loading mechanism and an inverted ring voltage, a
positron $g/2$ measurement would proceed identically to the electron measurement presented here.

Second, a direct measurement of the proton-to-electron mass ratio would combine the sub-ppb electron cyclotron
frequency resolution presented here with existing techniques for 90~ppt resolution of the proton cyclotron
frequency~\cite{FinalPbarMass} to compete with the existing 0.4~ppb limit~\cite{CODATA2006}.

Third, recent observations with a single trapped proton \cite{OneProtonSelfExcitedOscillator} open the way to proposed direct measurements of the proton and antiproton magnetic moments at the
ppb-scale~\cite{SelfExcitedOscillator,QuintAntiprotonAspirations}.  These would reduce the existing uncertainties by factors
of 10 and $10^{6}$, respectively, and provide an important test of CPT invariance~\cite{CptTestsInPenningTraps}. The
challenge to such a measurement is QND detection of a spin flip in a magnetic bottle because the smaller magnetic
moment and larger mass reduce the axial frequency shift by over $10^4$, though a larger bottle gradient can compensate
for some of the reduction.  Our goal is to realize with a proton and antiproton the great signal-to-noise ratio and
sensitivity to frequency changes that have been realized with the one-electron SEO.

Fourth, access to the lowest quantum states of a trapped electron and the lack of radiative damping of any degree of
freedom except cyclotron motion have led to several quantum information proposals using electrons in Penning traps,
e.g.,~\cite{Tombesi1999,Tombesi2001,Tombesi2003,Tombesi2005} that perhaps could be realized in a carefully optimized planar Penning trap \cite{OptimizedPlanarPenningTraps}.

\section{Conclusion}

In conclusion, precise control of the location of and the coupling to the electromagnetic modes of the electrode cavity
reduces the once-dominant cavity shift uncertainty.  This results in a measurement of the electron magnetic moment 15 times
more accurate than the 1987 measurement that provided the best $g/2$ and $\alpha$ for nearly 20 years.     With the measurement limited by the resolution and model of the
cyclotron and anomaly lines, future work on the electron $g/2$ should focus on enhancing magnetic field stability,
narrowing the lines, and building signal-to-noise. The techniques used in this result may be directly applied in
measuring the positron $g/2$ and may be adapted to a direct proton-to-electron mass ratio, to measurements of the
proton and antiproton magnetic moments, and to trapped-electron quantum information studies. With an independent $\alpha$ of
similar precision, the new $g/2$ would make possible 10-times more stringent tests of extensions to standard model.

%\section{Acknowledgements}
\begin{acknowledgments}
Many details about the experimental work are in a thesis \cite{ThesisHanneke}. This work was supported by
the NSF AMO division.
\end{acknowledgments}

%\bibliography{c:/users/gabrielse/jerry/shared/synchronize/ggrefs2010}
%\end{document}

%merlin.mbs 2010-03-15 4.21a (PWD, AO, DPC)

\end{document}